\documentclass[11pt]{article}

\usepackage[utf8]{inputenc}
\usepackage[english]{babel}
\usepackage{graphicx}         
\usepackage[pdftex, colorlinks=true, linkcolor=blue, citecolor=blue, urlcolor=black]{hyperref}
\usepackage{braket}
\usepackage{amsmath}
\usepackage{amssymb}
\usepackage{amsthm}
\usepackage{mathtools}
\usepackage{bbm}		
\usepackage{mathrsfs}
\usepackage{array}		
\usepackage{color}
\usepackage[usenames,dvipsnames]{xcolor}
\usepackage{caption}
\usepackage{enumitem}
\usepackage{subeqnarray}	
\usepackage{wrapfig}

\usepackage{tikz}
\usetikzlibrary{arrows, decorations.markings}

\allowdisplaybreaks

\numberwithin{equation}{section}

\theoremstyle{plain}\newtheorem{definition}{Definition}[section]

\newtheorem{proposition}[definition]{Proposition}
\newtheorem{cor}[definition]{Corollary}
\theoremstyle{remark}

\theoremstyle{plain}

\theoremstyle{plain}\newtheorem{assumption}{Assumption}

\theoremstyle{plain}\newtheorem{theorem}{Theorem}

\newcommand{\assit}[1]{\begin{enumerate}[label={(\alph*)}, ref={\theassumption\alph*}]{#1}\end{enumerate}}

\newcommand{\D}{\mathcal{D}}
\newcommand{\R}{\mathbb{R}}
\newcommand{\C}{\mathbb{C}}
\newcommand{\N}{\mathbb{N}}
\newcommand{\fH}{\mathfrak{H}}
\newcommand{\Fock}{\mathcal{F}}
\newcommand{\Number}{\mathcal{N}}
\newcommand{\vac}{|\Omega\rangle}
\newcommand{\id}{\mathbbm{1}}
\newcommand{\cJ}{\mathcal{J}}

\newcommand{\cL}{\mathcal{L}}

\newcommand{\cU}{\mathcal{U}}
\newcommand{\cQ}{\mathcal{Q}}

\newcommand{\fC}{\mathfrak{C}}

\let\textl\l
\renewcommand{\l}{\ell}
\renewcommand{\i}{\mathrm{i}}
\newcommand{\e}{\mathrm{e}}
\newcommand{\hc}{\mathrm{h.c.}}

\newcommand{\sym}{\mathrm{sym}}

\newcommand{\Tr}{\mathrm{Tr}}

\newcommand{\bPhi}{{\boldsymbol{\phi}}}

\newcommand{\bj}{\boldsymbol{j}}
\newcommand{\bk}{\boldsymbol{k}}

\newcommand{\bm}{\boldsymbol{m}}

\renewcommand{\tilde}[1]{\widetilde{#1}}

\newcommand{\lr}[1]{\left\langle #1 \right\rangle}
\newcommand{\norm}[1]{\lVert#1\rVert}
\newcommand{\onorm}[1]{\lVert#1\rVert_\mathrm{op}}

\renewcommand{\d}{\mathop{}\!\mathrm{d}}
\newcommand{\dx}{\d x}
\newcommand{\dy}{\d y}
\newcommand{\dz}{\d z}

\newcommand{\ds}{\d s}

\newcommand{\xn}{x^{(n)}}
\newcommand{\xk}{x^{(k)}}

\newcommand{\Ubar}{\overline{U}}
\newcommand{\Vbar}{\overline{V}}

\newcommand{\ad}{a^\dagger}
\newcommand{\Ad}{A^\dagger}

\newcommand\mydots{,\makebox[1em][c]{.\hfil.\hfil.},}
\newcommand\mycdots{\makebox[1em][c]{$\cdot$\hfil$\cdot$\hfil$\cdot$}}

\newcommand{\ls}{\lesssim}

\makeatletter
\DeclareFontFamily{OMX}{MnSymbolE}{}
\DeclareSymbolFont{MnLargeSymbols}{OMX}{MnSymbolE}{m}{n}
\SetSymbolFont{MnLargeSymbols}{bold}{OMX}{MnSymbolE}{b}{n}
\DeclareFontShape{OMX}{MnSymbolE}{m}{n}{
    <-6>  MnSymbolE5
   <6-7>  MnSymbolE6
   <7-8>  MnSymbolE7
   <8-9>  MnSymbolE8
   <9-10> MnSymbolE9
  <10-12> MnSymbolE10
  <12->   MnSymbolE12
}{}
\DeclareFontShape{OMX}{MnSymbolE}{b}{n}{
    <-6>  MnSymbolE-Bold5
   <6-7>  MnSymbolE-Bold6
   <7-8>  MnSymbolE-Bold7
   <8-9>  MnSymbolE-Bold8
   <9-10> MnSymbolE-Bold9
  <10-12> MnSymbolE-Bold10
  <12->   MnSymbolE-Bold12
}{}
\let\llangle\@undefined
\let\rrangle\@undefined
\DeclareMathDelimiter{\llangle}{\mathopen}
                     {MnLargeSymbols}{'164}{MnLargeSymbols}{'164}
\DeclareMathDelimiter{\rrangle}{\mathclose}
                     {MnLargeSymbols}{'171}{MnLargeSymbols}{'171}
\makeatother

\newcommand\smallO[1]{
        \mathchoice
            {
                \ensuremath{\mathop{}\mathopen{}{\scriptstyle\mathcal{O}}\mathopen{}\left(#1\right)}
            }
            {
                \ensuremath{\mathop{}\mathopen{}{\scriptstyle\mathcal{O}}\mathopen{}\left(#1\right)}
            }
            {
                \ensuremath{\mathop{}\mathopen{}{\scriptscriptstyle\mathcal{O}}\mathopen{}\left(#1\right)}
            }
            { 
                \ensuremath{\mathop{}\mathopen{}{o}\mathopen{}\left(#1\right)}
            }
    }

\newcommand{\fHN}{{\fH^N}}
\newcommand{\fHp}{\fH_{\perp\varphi}}

\newcommand{\trap}{\mathrm{trap}}
\newcommand{\free}{\mathrm{free}}

\newcommand{\HN}{H_N}
\newcommand{\HNtrap}{H_N^\mathrm{trap}}
\newcommand{\HNfree}{H_N^\mathrm{free}}
\newcommand{\Vext}{V^\mathrm{trap}}
\newcommand{\lN}{\lambda_N}

\newcommand{\PsiN}{\Psi_N}
\newcommand{\PsiNtrap}{\PsiN^\mathrm{trap}}
\newcommand{\PsiNfree}{\PsiN^\free}

\newcommand{\EN}{\mathscr{E}_N}
\newcommand{\ENtrap}{\EN^\mathrm{trap}}

\newcommand{\ENnu}{\mathscr{E}_N^{(\nu)}}

\newcommand{\pp}{p^\varphi}
\newcommand{\qp}{q^\varphi}

\newcommand{\goNtrap}{\gamma_N^{\trap,(1)}}
\newcommand{\goNfree}{\gamma_N^{\free,(1)}}
\newcommand{\goN}{\gamma_N^{(1)}}

\newcommand{\dnu}{\delta^{(\nu)}_N}
\newcommand{\dzn}{\delta^{(n)}_0}

\newcommand{\hH}{h}
\newcommand{\eH}{e_\mathrm{H}}
\newcommand{\mH}{\mu_\mathrm{H}}
\newcommand{\cEH}{\mathcal{E}_\mathrm{H}}

\newcommand{\UNp}{\mathfrak{U}_{N,\varphi}}

\newcommand{\Fp}{{\Fock_{\perp\varphi}}}
\newcommand{\FNp}{{\Fock_{\perp\varphi}^{\leq N}}}

\newcommand{\ChiN}{\Chi_{\leq N}}

\newcommand{\FockHN}{\FockH_{\leq N}}

\newcommand{\FockH}{\mathbb{H}}
\newcommand{\FockHz}{\FockH_0}
\newcommand{\FockHo}{\FockH_1}
\newcommand{\FockHt}{\FockH_2}

\newcommand{\FockHj}{\FockH_j}

\newcommand{\FockHjo}{\FockH_{j_1}}
\newcommand{\FockHjt}{\FockH_{j_2}}

\newcommand{\FockHjnu}{\FockH_{j_\nu}}

\newcommand{\tFockH}{\FockH^<}
\newcommand{\FockHminus}{\FockH^<}
\newcommand{\FockHplus}{\FockH^>}

\newcommand{\FockR}{\mathbb{R}}

\newcommand{\FockRnu}{\FockR_\nu}

\newcommand{\Chi}{{\boldsymbol{\chi}}}

\newcommand{\Chiz}{\Chi_0}

\renewcommand{\P}{\mathbb{P}}
\newcommand{\Pn}{\P}
\newcommand{\Q}{\mathbb{Q}}
\newcommand{\Qn}{\Q}

\newcommand{\Pz}{\P_0}
\newcommand{\Pzn}{\Pz}
\newcommand{\Qz}{\Q_0}
\newcommand{\Qzn}{\Qz}

\newcommand{\Ez}{E_0}

\newcommand{\Ezn}{\Ez}

\newcommand{\En}{E}

\newcommand{\In}{\iota^{(n)}}

\newcommand{\Np}{\Number_{\perp\varphi}}

\newcommand{\Ko}{K_1}
\newcommand{\Kt}{K_2}
\newcommand{\Kth}{K_3}
\newcommand{\Kf}{K_4}

\newcommand{\boldKz}{\mathbb{K}_0}
\newcommand{\boldKo}{\mathbb{K}_1}
\newcommand{\boldKt}{\mathbb{K}_2}
\newcommand{\boldKth}{\mathbb{K}_3}
\newcommand{\boldKf}{\mathbb{K}_4}
\newcommand{\boldKtbar}{\mathbb{K}_2^*}
\newcommand{\boldKthbar}{\mathbb{K}_3^*}

\newcommand{\UP}{\cU_P}

\newcommand{\Pnl}{\Pn_\l}

\newcommand{\ResHz}{\frac{1}{z-\FockH}}

\newcommand{\ResHzz}{\frac{1}{z-\FockHz}}

\newcommand{\FockO}{\mathbb{O}}

\newcommand{\FockB}{\mathbb{B}}
\newcommand{\FockBPn}{\FockB_P}
\newcommand{\FockBQn}{\FockB_Q}

\newcommand{\FockI}{\mathbb{I}}
\newcommand{\FockIn}{\FockI}
\newcommand{\FockInk}{\FockIn_k}

\newcommand{\gan}{\gamma}
\newcommand{\goint}{\oint_{\gan}}

\newcommand{\Am}{A^{(m)}}
\newcommand{\Amj}{\Am_{j_1\mydots j_m}}

\newcommand{\cAm}{\mathcal{A}^{(m)}_N}

\newcommand{\FockA}{\mathbb{A}}
\newcommand{\FockAmred}{\FockA^{(m)}_\mathrm{red}}

\newcommand{\BogU}{\mathbb{U}_\BogV}
\newcommand{\BogV}{\mathcal{V}}

\newcommand{\BogUz}{\mathbb{U}_\BogVz}
\newcommand{\BogVz}{{\mathcal{V}_0}}

\newcommand{\pt}{{\varphi(t)}}
\newcommand{\ps}{{\varphi(s)}}

\newcommand{\pz}{{\varphi(0)}}

\newcommand{\hpt}{h^{\pt}}
\newcommand{\mpt}{\mu^{\pt}}

\newcommand{\FN}{\Fock^{\leq N}}
\newcommand{\Fpz}{\Fock_{\perp\pz}}
\newcommand{\Fpt}{{\Fock_{\perp\pt}}}

\newcommand{\FNpt}{{\Fock^{\leq N}_{\perp\pt}}}

\newcommand{\UNpt}{\mathfrak{U}_{N,\pt}}
\newcommand{\UNpz}{\mathfrak{U}_{N,\pz}}

\newcommand{\Chil}{\Chi_\l}
\newcommand{\Chim}{\Chi_m}
\newcommand{\Chio}{\Chi_0}

\newcommand{\PsiNl}{\Psi_{N,\l}}

\newcommand{\FockHNpt}{\FockHN^{\pt}}
\newcommand{\FockHpt}{\FockH^{\pt}}
\newcommand{\FockHps}{\FockH^{\ps}}

\newcommand{\FockHopt}{\FockHpt_0}
\newcommand{\FockHnpt}{\FockHpt_n}
\newcommand{\FockHnps}{\FockHps_n}

\newcommand{\qpt}{q^{\pt}}
\newcommand{\Kopt}{K_1^{\pt}}

\newcommand{\Ktpt}{K_2^{\pt}}

\newcommand{\Kthpt}{K_3^{\pt}}

\newcommand{\BogUts}{\,\mathbb{U}_{\BogV(t,s)}}
\newcommand{\BogUtz}{\,\mathbb{U}_{\BogV(t,0)}}

\newcommand{\gChiot}{\gamma_{\Chio(t)}}
\newcommand{\gChioz}{\gamma_{\Chio(0)}}

\newcommand{\aChiot}{\alpha_{\Chio(t)}}
\newcommand{\aChioz}{\alpha_{\Chio(0)}}

\newcommand{\bzo}{\beta_{0,1}}

\newcommand{\ppt}{p^{\pt}}

\newcommand{\asjo}{a^{\sharp_{j_1}}}

\newcommand{\lrt}[1]{\left\langle #1 \right\rangle^{(t)}}

\title{Low-energy spectrum and dynamics of the\\  weakly interacting  Bose gas}
\author{Lea Boßmann\thanks{Institute of Science and Technology Austria, Am Campus 1, 3400 Klosterneuburg, Austria. \texttt{lea.bossmann@ist.ac.at}}}
\date{\today}

\begin{document}
\maketitle
%

\begin{abstract}
We consider a gas of $N$ bosons with interactions in the mean-field scaling regime.
We review the proof of an asymptotic expansion of its low-energy spectrum, eigenstates and dynamics, which provides corrections to Bogoliubov theory to all orders in $1/N$.
This is based on joint works with S.\ Petrat, P.\ Pickl, R.\ Seiringer and A.\ Soffer. 
In addition, we derive a full asymptotic expansion of the ground state one-body reduced density matrix.
\end{abstract}

\section{Introduction and main results}

\subsection{Introduction}
 
Since the first experimental realization of Bose--Einstein condensation (BEC) in 1995, the experimental, theoretical and mathematical investigation of systems of interacting bosons at low temperatures has become a very active field of research.
In a typical experiment, the bosons are initially caught in an external trap. This situation is mathematically described by the $N$-body Hamiltonian
\begin{equation}\label{HNtrap}
\HNtrap=\sum\limits_{j=1}^N\left(-\Delta_j+\Vext(x_j)\right)+\sum\limits_{1\leq i<j\leq N}v_N(x_i-x_j)
\end{equation}
for some confining  potential $\Vext$ and for some two-body interaction $v_N$, acting on the Hilbert space of square integrable, permutation symmetric functions on $\R^{dN}$,
\begin{equation*}
\fH^N_\sym:=\bigotimes\limits_\sym^N\fH\,,\qquad \fH:=L^2(\R^d)\,.
\end{equation*}
The Bose gas is then  cooled down to a low-energy eigenstate of $\HNtrap$, or to a superposition of such states. For simplicity, let us assume that the gas is prepared in the ground state $\PsiNtrap$ of $\HNtrap$, i.e.,
\begin{equation}\label{gs}
\ENtrap=\inf\sigma(\HNtrap)\,,\qquad \HNtrap\PsiNtrap= \ENtrap\PsiNtrap\,.
\end{equation}
Subsequently, the trap is switched off and the Bose gas propagates freely. Mathematically, this is described by the $N$-body Schrödinger equation with initial datum $\PsiNtrap$,
\begin{equation}\label{SE}
\i\partial_t \PsiNfree(t) = \HNfree \PsiNfree(t)\,,\qquad
\PsiNfree(0)=\PsiNtrap\,,
\end{equation}
with $N$-body Hamiltonian
\begin{equation}\label{HNfree}
\HNfree=\sum\limits_{j=1}^N\left(-\Delta_j\right)+\sum\limits_{1\leq i<j\leq N}v_N(x_i-x_j)\,.
\end{equation}
Given that the number of particles in such a gas is usually large, an exact (analytical or numerical) analysis of the system in presence of interactions is impossible. 
Over the last two decades, there have been many works in the  mathematical physics community devoted to a rigorous derivation of suitable approximations of the statical and dynamical properties of the gas for large $N$.
These questions have been studied for different classes of interactions $v_N$,
in particular for the so-called mean-field (or Hartree) regime
\begin{equation}\label{int}
v_N=\lN v\,,\qquad \lN:=\frac{1}{N-1}
\end{equation}
describing the situation of weak and long-range interactions.

In this note, we consider interactions of the form \eqref{int}.
We present an asymptotic expansion of the low-energy spectrum and eigenstates of $\HNtrap$ and of the dynamics \eqref{SE}, which makes the model fully computationally accessible to any order in $1/N$. This review is based on \cite{spectrum} (in collaboration with S.\ Petrat and R.\ Seiringer) and \cite{QF} (in collaboration with S.\ Petrat, P.\ Pickl and A.\ Soffer).

\subsection{Model and main results}

We consider a system of $N$ interacting bosons in $\R^d$, $d\geq1$, which are described by the $N$-body Hamiltonian \eqref{HNtrap} with interactions \eqref{int}. We impose the following assumptions on the interaction $v_N$ and the external potential $\Vext$:

\begin{assumption}\label{ass:v}
Define $v_N$ as in \eqref{int}. 
\assit{
\item Let $v:\R^d\to\R$ be bounded with $v(-x)=v(x)$ and $v\not\equiv 0$. \label{ass:v:bd}
\item Assume that $v$ is of positive type, i.e., that it has a non-negative Fourier transform.\label{ass:v:pos}
}
\end{assumption}

\begin{assumption}\label{ass:V}
Let $\Vext:\R^d\to\R$ be measurable, locally bounded and  non-negative and let $\Vext(x)$ tend to infinity as $|x|\to  \infty$.
\end{assumption}

Our first main result concerns the ground state $\PsiNtrap$ of $\HNtrap$: We construct a norm approximation of $\PsiNtrap$ and of its energy $\ENtrap$ to any order in $1/N$.

\begin{theorem}\label{thm:eigenstates:spectrum}
Let $a\in\N_0$, let Assumptions \ref{ass:v} and \ref{ass:V} be satisfied and choose $N$ sufficiently large. Then there exists a constant $C(a)$ such that
\begin{equation}\label{eqn:thm:eigenstates}
\Big\|\PsiNtrap-\sum\limits_{\l=0}^a \lN^\frac{\l}{2}\psi_{N,\l}^\trap\Big\|_{\fH^N}\leq C(a)\lN^\frac{a+1}{2}
\end{equation}
and
\begin{equation}\label{eqn:thm:spectrum}
\left|\ENtrap -  N\eH^\trap -  \sum\limits_{\l=0}^a\lN^\l E_\l^\trap \right|\; \leq\; C(a) \lN^{a+1}\,.
\end{equation}
The coefficients  $\psi_{N,\l}^\trap\in \fH_\sym^N$ of the expansion \eqref{eqn:thm:eigenstates} and  the coefficients
$\eH^\trap$, $E_\l^\trap \in\R$ of the expansion \eqref{eqn:thm:spectrum}
are given in \eqref{def:psi_N,l}, \eqref{def:eH} and \eqref{def:E_l}, respectively.
\end{theorem}
Our result extends to the low-energy excitation spectrum of $\HNtrap$ and to a certain class of unbounded interaction potentials $v$, including the repulsive three-dimensional Coulomb potential (see Section \ref{sec:spectrum:extensions}).
To leading order ($a=0$), the statements \eqref{eqn:thm:eigenstates} and \eqref{eqn:thm:spectrum}  have been proven (for bounded interactions) by Seiringer on the torus \cite{seiringer2011} and by Grech and Seiringer in the inhomogeneous setting \cite{grech2013}. For our class of unbounded interactions, the leading order approximation was obtained by Lewin, Nam, Serfaty and Solovej \cite{lewin2015_2}.  
The higher orders in \eqref{eqn:thm:eigenstates} and \eqref{eqn:thm:spectrum} were, to the best of our knowledge, first rigorously derived in \cite{spectrum}.
Another approach was proposed by Pizzo in \cite{pizzo2015,pizzo2015_2,pizzo2015_3}, who considers a Bose gas on a torus and constructs an expansion for the ground state, based on a multi-scale analysis in the number of excitations, around a product state using Feshbach maps.

As a consequence of the norm approximation \eqref{eqn:thm:eigenstates}, one can derive an expansion of the ground state one-body reduced density matrix,
\begin{equation}
\goNtrap:=\Tr_{\fH^{N-1}}|\PsiNtrap\rangle\langle\PsiNtrap|\,,
\end{equation}
in trace norm (see Section \ref{subsec:proof:RDM} for a proof of this statement):

\begin{cor}\label{cor:RDM:static}
Let $a\in\N_0$ and let Assumptions \ref{ass:v} and \ref{ass:V} be satisfied. 
Denote by $\gamma^{\trap,(1)}_N$ the one-body reduced density matrix of $\PsiNtrap$.
Then there exists a constant $C(a)>0$ such that
\begin{equation}\label{eqn:cor:RDM:static}
\Tr \Big| \goNtrap - \sum\limits_{\l=0}^a\lN^\l\gamma_{1,\l}^\trap \Big| \leq C(a) \lN^{a+1}
\end{equation}
for sufficiently large $N$, where the coefficients $\gamma_{1,\l}^\trap\in\cL(\fH)$ are defined in \eqref{eqn:exp:gamma:static}.
\end{cor}

Theorem \ref{thm:eigenstates:spectrum} and Corollary \ref{cor:RDM:static} determine the ground state $\PsiNtrap$ of $\HNtrap$ to arbitrary precision. 
Now we remove the confining potential $\Vext$ and take $\PsiNtrap$ as initial datum for the time evolution \eqref{SE}.
Since an eigenstate of $\HNtrap$ is not necessarily an eigenstate of $\HNfree$, this leads to some non-trivial dynamics, for which we provide an approximation in norm to any order in $1/N$ in our second main result:

\begin{theorem}\label{thm:dynamics}
Let $a\in\N_0$, $t\in\R$, let Assumption \ref{ass:v:bd} hold and denote by $\PsiNfree(t)$ the solution of \eqref{SE}.
Then there exists a constant $C(a)>0$ such that
\begin{equation}\label{eqn:thm:dynamics}
\Big\|\PsiNfree(t)-\sum\limits_{\l=0}^a\lN^{\frac{\l}{2}}\psi^\free_{N,\l}(t)\Big\|_{\fH^N}\leq \e^{C(a)t} \lN^{\frac{a+1}{2}}\,
\end{equation}
for sufficiently large $N$, where the coefficients $\psi^\free_{N,\l}(t)$ are defined in \eqref{eqn:PsiN_l}.
\end{theorem}
Note that for the dynamical result, we do not require the interaction potential to be of positive type.
Finally, we derive from the expansion \eqref{eqn:thm:dynamics} a trace norm approximation of the time-evolved reduced one-body reduced density matrix
\begin{equation}
\goNfree(t):=\Tr_{\fH^{N-1}}|\PsiNfree(t)\rangle\langle\PsiNfree(t)|
\end{equation}
to arbitrary precision:

\begin{cor}\label{cor:RDM:dynamics}
Let $a\in\N_0$, $t\in\R$ and let Assumption \ref{ass:v:bd} be satisfied. Then there exists a constant $C(a)$ such that
\begin{equation}\label{eqn:cor:RDM:dynamics}
\Tr \Big| \goNfree(t) - \sum\limits_{\l=0}^a\lN^\l\gamma_{1,\l}^\free(t) \Big| \leq \e^{C(a)t} \lN^{a+1}
\end{equation}
for sufficiently large $N$, where the coefficients $\gamma_{1,\l}^\free(t) \in\cL(\fH)$ are defined in \eqref{eqn:gamma:higher:orders}.
\end{cor}

Below, we will provide and explain the explicit formulas for the coefficients in Theorems \ref{thm:eigenstates:spectrum} and~\ref{thm:dynamics} and in Corollaries \ref{cor:RDM:static} and~\ref{cor:RDM:dynamics}. 
Note that $\eH^\trap$, $E^\trap_\l$, $\gamma_{1,\l}^\trap$ and $\gamma_{1,\l}^\free(t)$ are completely independent of $N$. The $N$-body wave functions $\psi_{N,\l}^\trap$ and $\psi_{N,\l}^\free(t)$ naturally depend on $N$; however, this $N$-dependence is  trivial, in a sense to be made precise below.
In particular, the computational effort to obtain physical quantities such as, e.g., expectation values of bounded operators with respect to the (time-evolved) $N$-body state, does not scale with $N$. 

Finally, let us remark that all constants $C(a)$ grow rapidly in $a$. Hence, all statements are to be read as \emph{asymptotic} expansions: given any order $a$ of the approximation, one can choose $N$ sufficiently large such that the estimates are meaningful, but we cannot simultaneously send $a$ to infinity.
\\

These notes are organized as follows: In Section \ref{sec:spectrum}, we explain the results from \cite{spectrum} concerning the low-energy spectrum and eigenstates and give a proof of Corollary \ref{cor:RDM:static}. Section~\ref{sec:dynamics} contains the results for the dynamics obtained in \cite{QF}.

\subsection*{Notation}
\begin{itemize}
\item 
The notation $A\ls B $
indicates that there exists a constant $C>0$ such that $A\leq CB$.
\item For $k\geq 1$ and $x_j\in{\R^d}$, we abbreviate
$\xk:=(x_1\mydots x_k)$ and $\d\xk:=\dx_1\mycdots\dx_k $.
\item We use the notation
$a^{\sharp_1}:=\ad$ and $ a^{\sharp_{-1}}:=a$.
\item
Multi-indices are denoted as $\bj=(j_1\mydots j_n)$ with $|\bj|:=j_1+\dots+j_n$.
\end{itemize}

\section{Low-energy spectrum and eigenstates}\label{sec:spectrum}

In this section, we consider the Hamiltonian $\HNtrap$ from \eqref{HNtrap} and explain the asymptotic expansion of its ground state $\PsiNtrap$, the ground state energy $\ENtrap$, and the corresponding reduced density matrix $\goNtrap$. To keep the notation simple, we drop the superscript $^\trap$.

\subsection{Framework}
\subsubsection{Condensate}

It is well known (see, e.g., \cite{seiringer2011, grech2013, lewin2014, lewin2015_2}) that the $N$-body ground state $\PsiN$ exhibits (complete asymptotic) BEC in the minimizer $\varphi\in\fH$ of the Hartree energy functional $\cEH$,
\begin{equation}\label{def:Hartree:functional}
\cEH[\phi]:=\int\limits_{\R^d}\left(|\nabla \phi(x)|^2+V(x)|\phi(x)|^2\right)\dx
+\tfrac12\int\limits_{\R^{2d}}v(x-y)|\phi(x)|^2|\phi(y)|^2\dx\dy\,.
\end{equation}
For potentials $v$ and $V$ satisfying Assumptions \ref{ass:v} and \ref{ass:V}, the minimizer $\varphi$ of $\cEH$ is unique, strictly positive and solves the stationary Hartree equation
\begin{equation}\label{def:h}
\hH\varphi:=(-\Delta+V+v*\varphi^2-\mH)\varphi
=0
\end{equation}
with Lagrange parameter
$
\mH:=\lr{\varphi,\left(-\Delta+V+v*\varphi^2\right)\varphi}\in\R\,.
$
We denote by $\pp$ and $\qp$ the projector onto $\varphi$ and its orthogonal complement, i.e., 
\begin{equation}\label{def:pp}
\pp:=|\varphi\rangle\langle\varphi|\,,\quad \qp:=\id-\pp\,.
\end{equation}
The minimum of $\cEH$ is given as
\begin{equation}\label{def:eH}
\eH:=\cEH[\varphi]=\lr{\varphi,\left(-\Delta+V+\tfrac12v*\varphi^2\right)\varphi}\,.
\end{equation}
Heuristically, (complete asymptotic) BEC in the state $\varphi$ means that $N-\smallO{N}$ particles occupy the condensate state $\varphi$.
Mathematically, this is reflected by the fact that the $N$-body wave function is determined by the one-body state $\varphi$ in the sense of reduced densities, i.e., 
\begin{equation}\label{eqn:BEC}
\lim\limits_{N\to\infty} \Tr\,\left|\goN-|\varphi\rangle\langle\varphi|\right| =0\,.
\end{equation}
The condensate determines the leading order of the ground state energy, namely
\begin{equation}\label{eqn:leading:order:gs:energy}
\EN= N\eH+\mathcal{O}(1)\,.
\end{equation}

\subsubsection{Excitations}
The errors in \eqref{eqn:BEC} and \eqref{eqn:leading:order:gs:energy} are caused by $\mathcal{O}(1)$ particles  which are excited from the condensate due to the inter-particle interactions.
To describe these excitations, we decompose $\PsiN$ as
\begin{equation}\label{eqn:decomposition:PsiN}
\PsiN=\sum\limits_{k=0}^N{\varphi}^{\otimes (N-k)}\otimes_s\chi^{(k)}\,,
\qquad \chi^{(k)}\in \bigotimes\limits_\sym^k \fHp\,, 
\end{equation}
with $\otimes_s$ the symmetric tensor product and where $\fHp :=\left\{\phi\in \fH: \lr{\phi,\varphi}_{\fH}=0\right\}$ denotes the orthogonal complement of $\varphi$ in $\fH$ \cite{lewin2015_2}.  
The excitations
\begin{equation}
\ChiN:=\big(\chi^{(k)}\big)_{k=0}^N
\end{equation}
form a vector in the truncated (excitation) Fock space over $\fHp$,
\begin{equation}\label{Fock:space}
\FNp=\bigoplus_{k=0}^N\bigotimes_\sym^k \fHp
\;\subset \;
\Fp=\bigoplus_{k=0}^\infty\bigotimes_\sym^k \fHp
\;\subset \;\Fock=\bigoplus_{k\geq 0}\bigotimes_\sym^k \fH\,,
\end{equation}
which is a subspace of the Fock space $\Fock$ over $\fH$. 
The creation/annihilation operators $\ad$/$a$ on $\Fock$ are defined in the usual way, and we denote the second quantization in $\Fock$  of an operator $T$ on $\fH$ by $\d\Gamma(T)$.
The number operator on $\Fp$ is given by
\begin{equation}
\Np:=\d\Gamma(\qp)\,.
\end{equation}
The relation between $\PsiN$ and the corresponding excitation vector $\ChiN$ is given by the unitary map
\begin{eqnarray}\label{map:U}
\UNp:\fHN \to  \FNp\;, \quad
 \Psi  \mapsto  \UNp \Psi= \ChiN\,,
\end{eqnarray}
whose action is explicitly known (see \cite[Proposition 4.2]{lewin2015_2}). 
Conjugating $\HN$ with $\UNp$ yields the operator 
\begin{equation}
\FockHN\;:=\;\UNp\left(\HN-N\eH\right)\UNp^*
\end{equation}
on $\FNp$, whose ground state is given by $\ChiN$. Hence, the ground state energy of $\FockHN$,
\begin{equation}
E_{\leq N}:=\inf\sigma(\FockHN)=\lr{\ChiN,\FockHN\ChiN}_{\FNp}=\EN-N\eH\,,
\end{equation} 
is precisely the $\mathcal{O}(1)$-term in \eqref{eqn:leading:order:gs:energy}.

\subsubsection{Excitation Hamiltonian}
Making use of the explicit form of $\UNp$ \cite[Proposition 4.2]{lewin2015_2}, we can express $\FockHN$ as
\begin{eqnarray}
\FockHN
&=&  \boldKz + \left(\frac{N-\Np}{N-1}\right) \boldKo\nonumber\\
&&+\left( \boldKt\frac{\sqrt{(N-\Np)(N-\Np-1)}}{N-1}+ \frac{\sqrt{(N-\Np)(N-\Np-1)}}{N-1} \boldKtbar\right)\nonumber\\
&&+\left( \boldKth\frac{\sqrt{N-\Np}}{N-1}+ \frac{\sqrt{N-\Np}}{N-1} \boldKthbar\right) 
+\frac{1}{N-1} \boldKf\,\label{eqn:FockHN}
\end{eqnarray}
as an operator on $\FNp$, where we used the shorthand notation
\begin{subequations}\label{eqn:K:notation}
\begin{eqnarray}
 \boldKz&:=&\d\Gamma(h)\,,\qquad \boldKo\;:=\;\d\Gamma(\Ko)\,, \qquad\boldKf\;:=\;\d\Gamma(\Kf)\,,
 \label{eqn:K:notation:0}\\
 \boldKt&:=&\tfrac12\int\dx_1\dx_2\, \Kt(x_1,x_2)\ad_{x_1}\ad_{x_2}\,,\label{eqn:K:notation:2}\\
 \boldKth&:=&\int\dx^{(3)}\, \Kth(x_1,x_2;x_3)\ad_{x_1}\ad_{x_2}a_{x_3}\,\label{eqn:K:notation:3} 
\end{eqnarray}
\end{subequations}
for $h$ as in \eqref{def:h} and where
\begin{subequations}\label{K}
\begin{align}
\label{K:1}
& \Ko:\fHp\to \fHp, \qquad
 \Ko :=   \qp   K  \qp \,, 
 \\[7pt]
\label{K:2}
& \Kt \in \fHp\otimes \fHp,\qquad
 \Kt(x_1,x_2) := 
   (\qp_1\qp_2 K)(x_1,x_2)\,,
\\[7pt]
\label{K:3}
& \Kth: \fHp \to \fHp\otimes \fHp, \;\;\;
(\Kth\psi)(x_1,x_2):=\qp_1\qp_2 W(x_1,x_2)\varphi(x_1)(\qp_2\psi)(x_2)
\,,  \\[3pt]
\label{K:4}
& \Kf:\fHp\otimes \fHp\to\fHp\otimes \fHp,\;\;\;
(\Kf\psi)(x_1,x_2):=\qp_1\qp_2W(x_1,x_2)(\qp_1\qp_2\psi)(x_1,x_2)\,.
\end{align}
\end{subequations}
Here, $K(x_1,x_2)$ is defined as 
\begin{equation}\label{def:K:kernel}
 K(x_1;x_2):=v(x_1-x_2)\varphi(x_1)\varphi(x_2)\,,
\end{equation}
$K$  is the operator with kernel $K(x_1,x_2)$, and $W$ is the multiplication operator defined by
\begin{equation}\label{eqn:W(x,y)}
W(x_1,x_2) := v(x_1-x_2) - \left(v*\varphi^2\right)(x_1) - \left(v*\varphi^2\right)(x_2) + \lr{\varphi,v*\varphi^2\varphi}.
\end{equation}
\medskip

By construction, $\FockHN$ is explicitly $N$-dependent. 
To extract its contributions to each order in $\lN$, we first extend $\FockHN$ trivially to an operator on $\Fp$,
\begin{equation}\label{def:FockH}
\FockH:=\FockHN\oplus c\,,
\end{equation}
where the direct sum is with respect to the decomposition $\Fock=\Fock^{\leq N}\oplus\Fock^{<N}$. The constant $c$ in \eqref{def:FockH} will later be chosen conveniently (see Section \ref{sec:spectrum:proof}). Similarly,  we extend $\ChiN$ to a vector $\Chi\in\Fp$ as
\begin{equation}
\Chi:=\ChiN\oplus 0
\end{equation} 
and denote the corresponding projectors on $\Fp$ by
\begin{equation}\label{def:P}
\P:=|\Chi\rangle\langle\Chi|\,,\qquad \Q:=\id-\P\,.
\end{equation}
A (formal) expansion of $\FockH$ in powers of $\lN^{1/2}$ yields
\begin{equation}\label{intro:taylor}
\FockH \;=\;\FockHz+\sum\limits_{j\geq 1}\lN^\frac{j}{2}\FockHj\,,
\end{equation}
where 
\begin{subequations}\label{FockHj}
\begin{eqnarray}
\FockHz&:=&\boldKz+\boldKo+\boldKt+\boldKtbar\,, \\[5pt]
\FockHo&:=& \boldKth + \boldKthbar\,,\label{FockHo}\\[5pt]
\FockHt
&:=& -(\Np-1)\boldKo -\Big(\boldKt(\Np-\tfrac12)+\hc\Big)+\boldKf\,,\label{FockHt}\\[5pt]
\FockH_{2j-1}
&:=&c_{j-1}\Big(\boldKth (\Np-1)^{j-1} + \hc\Big)\,,\label{FockHt:2n-1}\\
\FockH_{2j}&:=&
\sum\limits_{\nu=0}^j d_{j,\nu}\Big(\boldKt(\Np-1)^\nu +\hc\Big)\,\label{FockHt:2n}
\end{eqnarray}
\end{subequations}
for $j\geq 2$, with $\mathbb{K}_j$ as in \eqref{eqn:K:notation}. The coefficients $c_j$ and $d_{j,\nu}$ are given as
\begin{subequations}\label{taylor:coeff}
\begin{eqnarray}\label{eqn:taylor:coeff}
c^{(\l)}_0&:=&1\,,\quad c^{(\l)}_j \;:=\; \frac{(\l-\frac12)(\l+\frac12)(\l+\frac32)\mycdots(\l+j-\frac32)}{j!}\,,\quad c_j\;:=\;c_j^{(0)} \quad (j\geq1),\qquad\\
\label{eqn:taylor:coeff:2}
d_{j,\nu}&:=&\sum\limits_{\l=0}^\nu c_\l^{(0)} c_{\nu-\l}^{(0)} c_{j-\nu}^{(\l)} \qquad (j\geq \nu \geq 0)\,.
\end{eqnarray}
\end{subequations}

\subsubsection{Bogoliubov approximation}

The leading order term $\FockHz$ in \eqref{intro:taylor} is the well-known Bogoliubov Hamiltonian.
We denote the unique ground state of $\FockHz$ and the ground state energy by
\begin{equation}\label{def:Chiz}
\Ez:=\inf\sigma(\FockHz)\,,\qquad \FockHz\Chiz=\Ez\Chiz\,,
\end{equation}
and the corresponding projectors are defined as
\begin{equation}\label{def:Pz}
\Pz:=|\Chiz\rangle\langle\Chiz|\,, \qquad \Qz:=\id-\Pz\,.
\end{equation}
It is well known \cite{seiringer2011,grech2013,lewin2015_2} that the ground state $\ChiN$ of $\FockHN$ and the ground state energy $E_{\leq N}$ converge to $\Chiz$ and $\Ez$, respectively, i.e.,
\begin{equation}\label{eqn:known:results:FockHN:E}
\lim\limits_{N\to\infty}E_{\leq N}=\Ez\,,\qquad
\lim\limits_{N\to\infty}\norm{\ChiN-\Chiz}_{\FNp}=0\,.
\end{equation}
Consequently, $\Ez$ gives the next-to-leading order term in \eqref{eqn:thm:spectrum}; analogously, the leading order contribution in \eqref{eqn:thm:eigenstates} is given by $\psi_{N,0}=\UNp^*\,\Chiz|_{\FNp}$.
\medskip

The Bogoliubov Hamiltonian $\FockHz$ is a very useful approximation of $\FockH$ because it is much simpler than the full problem: it is quadratic in the number of creation/annihilation operators and can be diagonalized by Bogoliubov transformations.

Let us briefly recall the concept of Bogoliubov transformations.
For $F=f\oplus Jg \in\fH\oplus\fH$, where  $J:\fH\to\fH$ denotes complex conjugation, one defines the generalized creation and annihilation operators $A(F)$ and $\Ad(F)$ as
\begin{equation}\label{eqn:A(F)}
A(F)=a(f)+\ad(g)\,, \quad \Ad(F)=A(\cJ F)=\ad(f)+a(g)
\end{equation}
for
$ \cJ=\left(\begin{smallmatrix}0 & J\\J&0\end{smallmatrix}\right)$. 
An operator $\BogV$ on $\fH\oplus\fH$  such that  $F\mapsto A(\BogV F)$ has the same properties as  $F\mapsto A(F)$, i.e., 
$
\Ad(\BogV F)=A(\BogV\mathcal{J}F)$ and $[A(\BogV F_1),\Ad(\BogV F_2)]=[A(F_1),\Ad(F_2)]\,,
$
is called a \emph{(bosonic) Bogoliubov map} and can be written in block form as
\begin{equation}\label{BogV:block:form}
\BogV:=\begin{pmatrix}U & \Vbar\\V & \Ubar\end{pmatrix}\,,\quad U,V:\fHp\to \fHp\,.
\end{equation}
If  $V$ is Hilbert-Schmidt, the Bogoliubov map $\BogV$ can be unitarily implemented on $\Fock$, i.e., there exists a unitary transformation $\BogU:\Fock\to\Fock$ (called a \emph{Bogoliubov transformation}) such that
$
\BogU A(F)\BogU^*=A(\BogV F)
$
for all  $F\in\fH\oplus\fH$. This implies the transformation rule
\begin{equation}\begin{split}
\BogU \,a(f)\,\BogU^*&\;=\;a(Uf) +\ad (\overline{V f})
\,,\qquad
\BogU\,\ad(f)\,\BogU^*\;=\;\ad(Uf) +a (\overline{V f})\,.\label{eqn:trafo:ax}
\end{split}\end{equation}
A normalized state $\bPhi\in\Fp$ which can be written as
\begin{equation}
\bPhi=\BogU|\Omega\rangle
\end{equation}
for some Bogoliubov map $\BogV$ is called a \emph{quasi-free  state}. Quasi-free states have a finite expectation value of the number operator and satisfy Wick's rule, i.e.,
\begin{subequations}\label{eqn:Wick}
\begin{align}
\lr{\bPhi,a^\sharp(f_1)\mycdots a^\sharp(f_{2n-1})\bPhi}_{\Fp} &= 0\,,\\
\lr{\bPhi,a^\sharp(f_1)\mycdots a^\sharp(f_{2n})\bPhi}_{\Fp} &= \sum\limits_{\sigma\in P_{2n}}\prod\limits_{j=1}^n \lr{\bPhi,a^\sharp(f_{\sigma(2j-1)})a^\sharp(f_{\sigma(2j)})\bPhi}_{\Fp}
\end{align}
\end{subequations}
for $a^\sharp\in\{\ad,a\}$, $n\in\N$ and $f_1\mydots f_{2n}\in \fHp$.
Here, $P_{2n}$ denotes the set of pairings
\begin{equation}\label{pairings}
P_{2n}:=\{\sigma\in\mathfrak{S}_{2n}:\sigma(2a-1)<\min\{\sigma(2a),\sigma(2a+1)\} \;\forall a\in \{1,2\mydots 2n\} \}\,,
\end{equation}
for $\mathfrak{S}_{2n}$ the symmetric group on the set $\{1,2\mydots 2n\}$.
In particular, the ground state $\Chiz$ of $\FockHz$ is a quasi-free state, 
\begin{equation}
\Chiz=\BogUz^*\vac\,,
\end{equation}
where $\BogUz$ is the Bogoliubov transformation that diagonalizes $\FockHz$.

\subsection{Expansion of the ground state}
To prove Theorem \ref{thm:eigenstates:spectrum}, we show that the projector $\P$ from \eqref{def:P} admits a series expansion in powers of $\lN^{1/2}$ in the following sense:
\begin{proposition}\label{prop:static:technical}
Let  Assumptions \ref{ass:v} and \ref{ass:V} hold, let $\FockA\in\cL(\Fp)$ be a bounded operator on $\Fp$ and let $a\in\mathbb{N}_0$. Then there exists some constant $C(a)$ such that
\begin{equation}
\left|\Tr\,\FockA\P-\sum\limits_{\l=0}^a\lN^\frac{\l}{2}\Tr\,\FockA\P_\l\right|\leq C(a) \lN^\frac{a+1}{2} \onorm{\FockA}
\end{equation}
for sufficiently large $N$, where $\onorm{\cdot}$ denotes the operator norm. The coefficients $\P_\l$  are defined as
\begin{equation}\label{eqn:Pna}
\P_\l:=\begin{cases}
\quad \Pz & \text{ if }\; \l=0\,,\\[5pt]
\displaystyle -\sum\limits_{\nu=1}^\l\,
\sum\limits_{\substack{\bj\in\N^\nu\\[2pt]|\bj|=\l}} \,
\sum\limits_{\substack{\bk\in\N_0^{\nu+1}\\|\bk|=\nu}} 
\FockO_{k_1}\FockHjo\FockO_{k_2}\FockHjt
\mycdots
\FockO_{k_\nu} \FockHjnu\FockO_{k_{\nu+1}} & \text{ if }\;\l\geq 1\,,
\end{cases}
\end{equation}
with $\Pz$ as in \eqref{def:Pz} and $\FockH_j$ as in \eqref{FockHj} and where we abbreviated
\begin{equation}\label{FockO}
\FockO_k:=\begin{cases}
\displaystyle\quad-\Pz &\quad k=0\,,\\[7pt]
\displaystyle\frac{\Qz}{\big(\Ez-\FockHz\big)^k} &\quad k>0\,.
\end{cases}
\end{equation}
\end{proposition}
The growth of the constant $C(a)$ in the order $a$ of the approximation can be estimated as
\begin{equation*}
C(a) \leq C (a+1)^{(a+6)^2} \,,
\end{equation*} 
which we expect to be far from optimal.
By means of Bogoliubov transformations, the operators $\Pnl$  can be brought into a more explicit form. For example, the first order correction $\P_1$ is given by
\begin{equation}\label{eqn:Pl:explicit:1}
\P_1 = \BogUz^*\left(\BogUz\FockO_1\BogUz^*\right)\left(\BogUz\FockH_1\BogUz^*\right)\vac\langle\Chiz|+\hc\,,
\end{equation}
where $\BogUz$ is the Bogoliubov transformation diagonalizing $\FockHz$ such that $\Chiz=\BogUz^*\vac$. 
To simplify \eqref{eqn:Pl:explicit:1}, one notes that $\BogUz\FockH_1\BogUz^*\vac$ is a superposition of one- and three-particle states and that
$\BogUz\FockO^{(0)}_1\BogUz^*$ is particle-number preserving.
Hence, $\P_1$ can be expressed as
\begin{equation}\label{eqn:Theta:1}
\P_1=\BogUz^*\left(\int\dx\,\Theta_1(x)\ad_x|\Omega\rangle + \int\dx^{(3)}\Theta_3(x^{(3)})\ad_{x_1}\ad_{x_2}\ad_{x_3}|\Omega\rangle\right)\langle\Chiz|+\hc\,,
\end{equation}
where the functions $\Theta_1$ and $\Theta_3$ can be retrieved by diagonalizing $\FockHz$ and computing the Bogoliubov transformation of $\FockH_1$ under $\BogUz$.
\\

From Proposition \ref{prop:static:technical}, we deduce three consequences:

\subsubsection{Ground state wave function}
As an immediate consequence of Proposition \ref{prop:static:technical}, we find that
\begin{equation}\label{eqn:trace:norm}
\Tr\,\Big|\P-\sum\limits_{\l=0}^a\lN^\frac{\l}{2}\P_\l\Big| \leq C(a)\lN^\frac{a+1}{2}\,.
\end{equation}
Since $\P=|\Chi\rangle\langle\Chi|$ is a rank one projector, the expansion \eqref{eqn:trace:norm} implies an expansion of the excitation wave function $\Chi$,
\begin{equation}\label{eqn:expansion:Chi}
\Big\|\Chi-\sum\limits_{\l=0}^a\lN^\frac{\l}{2}\Chi_\l\Big\|_\Fock \leq C(a) \lN^\frac{a+1}{2}
\end{equation} 
(see \cite[Appendix B]{spectrum} for a proof of this statement in a general Hilbert space setting). The coefficients of the expansion are given by
\begin{equation}
\Chi_\l:=\sum\limits_{j=0}^\l \alpha_j\,\tilde{\Chi}_{\l-j} \qquad (\l\geq 1)\label{chi_def_thm}
\end{equation}
where
\begin{equation}
\tilde{\Chi}_\l:=\sum\limits_{\nu=1}^\l\sum\limits_{\substack{\bj\in\N^\nu\\|\bj|=\l}} \P_{j_1}\,\mycdots \P_{j_\nu}\,\Chi_0 \qquad (\l\geq 1)\,,\label{chi_tilde_def_thm}
\end{equation}
with $\P_\l$ as in \eqref{eqn:Pna}, $\Chiz$ as in \eqref{def:Chiz} and, for $n\geq 1$,
\begin{equation}
\alpha_0:=1\,,\qquad
\alpha_{2n-1}:=0\,,\qquad
\alpha_{2n}:=-\frac12\sum\limits_{\substack{\bj\in\N_0^4\\j_1,j_2<2n\\|\bj|=2n}} \alpha_{j_1}\alpha_{j_2}\lr{\tilde{\Chi}_{j_3},\tilde{\Chi}_{j_4}}\,. 
\end{equation}
For example, 
\begin{equation}\label{eqn:Theta:2}
\Chi_1=\frac{\Qz}{\Ez-\FockHz}\FockH_1\Chiz = \BogUz^*\left(\int\dx\,\Theta_1(x)\ad_x|\Omega\rangle + \int\dx^{(3)}\Theta_3(x^{(3)})\ad_{x_1}\ad_{x_2}\ad_{x_3}|\Omega\rangle\right)
\end{equation}
for $\Theta_1$ and $\Theta_3$ as in \eqref{eqn:Theta:1}.
Finally, the coefficients $\psi_{N,\l}$ in the expansion \eqref{eqn:thm:eigenstates} of the $N$-body ground state $\PsiN$  (Theorem \ref{thm:eigenstates:spectrum}) are constructed from this by inserting \eqref{chi_def_thm} into \eqref{eqn:decomposition:PsiN}, i.e.,
\begin{equation}\label{def:psi_N,l}
\psi_{N,\l} := \sum\limits_{k=0}^N\varphi^{\otimes(N-k)}\otimes_s(\Chi_\l)^{(k)}\,.
\end{equation}
The functions $\psi_{N,\l}$ depend on $N$ by construction. However, this $N$-dependence is trivial, since it comes only from the splitting into condensate $\varphi$ and excitations $\Chi$. The coefficients $\Chi_\l$ in the expansion \eqref{eqn:thm:eigenstates} of the  excitations $\Chi$ are completely independent of $N$.

\subsubsection{Ground state energy}
Another consequence of Proposition \ref{prop:static:technical} is the expansion \eqref{eqn:thm:spectrum} of the ground state energy $\mathcal{E}_N$ (Theorem \ref{thm:eigenstates:spectrum}).
The coefficients $E_\l$ in \eqref{eqn:thm:spectrum} are given as
\begin{equation}\label{def:E_l}
E_\l\;:=\;
\sum\limits_{\nu=1}^{2\l}\sum\limits_{\substack{\bj\in\N^\nu\\|\bj|=2\l}}
\sum\limits_{\substack{\bm\in\N_0^{\nu-1}\\|\bm|=\nu-1}}\frac{1}{\kappa(\bm)}
\Tr\,\Pz\FockHjo\FockO_{m_1}\mycdots\FockH_{j_{\nu-1}}\FockO_{m_{\nu-1}}\FockHjnu
\end{equation}
for $\Pz$ as in \eqref{def:Pz}, $\FockH_j$ as in \eqref{FockHj}, $\FockO_m$ as in \eqref{FockO},  and where 
\begin{equation}
\kappa(\bm):=1+\left|\left\{\mu\,:\, m_\mu=0\right\}\right| \in\{1\mydots \nu-1\}
\end{equation} 
is the number of operators $\Pz$ within the trace.
This confirms the predictions of (formal) Rayleigh--Schrödinger perturbation theory. For example, the first coefficient in \eqref{def:E_l} simplifies  to
\begin{eqnarray}\label{eqn:cor:explicit}
E_1&=& \lr{\Chiz,\FockHt\Chiz} +\lr{\Chiz,\FockHo\frac{\Qzn}{\Ezn-\FockHz}\FockHo\Chiz}
\,.
\end{eqnarray}

\subsubsection{Ground state reduced density}

Finally, Proposition \ref{prop:static:technical} implies an asymptotic expansion of the one-body reduced density $\gamma_N^{(1)}$ of $\PsiN$ (Corollary \ref{cor:RDM:static}). The coefficients in \eqref{eqn:cor:RDM:static} are given by the trace class operators with kernels
\begin{subequations}\label{eqn:exp:gamma:static}
\begin{eqnarray}
\gamma_{1,0}(x;y)&:=&\varphi(x)\varphi(y)\,,\\
\gamma_{1,\l}(x;y)&:=&\sum\limits_{n=0}^{\l-1}\sum\limits_{k=0}^{\l-n-1}\tilde{c}_{\l-n-1,k}
\left(\varphi(x)\Tr\,\P_{2n+1}\ad_y(\Np-1)^k +  \varphi(y)\Tr\,\P_{2n+1}(\Np-1)^ka_x \right)\nonumber\\
&&+\sum\limits_{n=0}^{\l-1}\tilde{c}_{\l-n-1}\left(\Tr\,\P_{2n}\ad_ya_x-\varphi(x)\varphi(y)\Tr\,\P_{2n}\Np)\right)
\end{eqnarray}
\end{subequations}
with 
\begin{equation}\label{def:c:tilde}
\tilde{c}_\l:=(-1)^\l c_\l^{(3/2)}\,,\qquad \tilde{c}_{\l,k}:=\tilde{c}_{\l-k}c_k^{(0)}
\end{equation} 
for $c_j^{(n)}$ as in \eqref{eqn:taylor:coeff}.
For example, the leading order is  $\gamma_{1,0}=\pp$, which recovers the well-known fact that the ground state exhibits BEC with optimal rate. The first correction to this is given by
\begin{equation}\label{eqn:1:p:RDM}\begin{split}
\gamma_{1,1} (x;y)\;=\;& \varphi(x)\Tr\,\Pn_1 \ad_y  + \varphi(y)\Tr\,\Pn_1 a_x \\
&+ \Tr\,\Pzn  \ad_y a_x -\varphi(x)\varphi(y)\Tr\,\Pzn\Np \,.
\end{split}\end{equation}
For the ground state of a homogeneous Bose gas on the torus, $\gamma_{1,1}$ was recently derived in \cite{nam2020_2}, using different methods. In that case, the first line in \eqref{eqn:1:p:RDM} vanishes by translation invariance. 
We prove Corollary \ref{cor:RDM:static} in Section \ref{subsec:proof:RDM}.

\subsection{Strategy of proof}\label{sec:spectrum:proof}
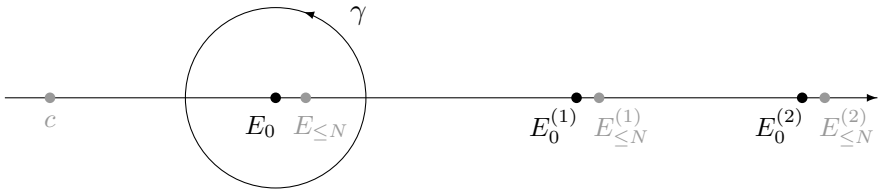
\begin{figure}[h]
\centering
\begin{tikzpicture}
\draw [decoration={ markings,
      mark=at position 1.0 with {\arrow{latex}}},postaction={decorate}] (-2.6,0)--(9,0); 
      
\fill[gray!80] (-2,0) circle (2pt);
\node at (-2,-0.1) [below] {\textcolor{gray!80}{\small $c$}};
      
\fill (1,0) circle (2pt);
\node at (0.8,-0.1) [below] {{\small $E_0$}};

\fill[gray!80] (1.4,0) circle (2pt);
\node at (1.6,-0.1) [below] {\textcolor{gray!80}{\small $E_{\leq N}$}};

\fill (5,0) circle (2pt);
\node at (4.7,0) [below] {{\small $E_0^{(1)}$}};

\fill[gray!80] (5.3,0) circle (2pt);
\node at (5.6,0) [below] {\textcolor{gray!80}{\small $E_{\leq N}^{(1)}$}};

\fill (8,0) circle (2pt);
\node at (7.7,0) [below] {{\small $E_0^{(2)}$}};

\fill[gray!80] (8.3,0) circle (2pt);
\node at (8.6,0) [below] {\textcolor{gray!80}{\small $E_{\leq N}^{(2)}$}};

\draw [decoration={ markings,
      mark=at position 0.2 with {\arrow{latex}}},postaction={decorate}] (1,0) circle (1.2cm);
\node at (2.1,1.1) {$\gamma$};
\end{tikzpicture}
  \caption{Low-energy spectra of $\FockHz$ (drawn in black) and $\FockH$ (drawn in grey). The additional eigenvalue $c$ of $\FockH$ is placed a finite distance below $E_0$. For sufficiently large $N$, the contour $\gamma$ around $E_0$ encloses the ground state energy $E_{\leq N}$ of $\FockHN$.}
  \label{fig}
\end{figure}

The first step is to express $\P$ and $\Pz$ as contour integrals around the resolvents of $\FockH$ and $\FockHz$, respectively, i.e.,
\begin{equation} \label{eqn:functional:calculus}
\P=\frac{1}{2\pi\i}\oint_\gamma\ResHz\dz\,,\qquad \Pz=\frac{1}{2\pi\i}\oint_\gamma\ResHzz\dz\,.
\end{equation}
The contour $\gamma$ is chosen such that its length is $\mathcal{O}(1)$ and that it encloses both the ground state energy $E_{\leq N}$ of $\FockHN$ and the Bogoliubov ground state energy $\Ez$ but leaves the remaining spectra of $\FockH$ and $\FockHz$ outside.  Since $E_{\leq N}$ converges to $\Ez$ as $N\to\infty$ by \eqref{eqn:known:results:FockHN:E}, such a contour exists if the constant $c$ in $\FockH=\FockHN\oplus c$ from \eqref{def:FockH} is chosen a finite distance away from the spectrum of $\FockHz$. This implies that $\FockH$ has precisely one (infinitely degenerate) additional eigenvalue $c$ compared to $\FockHN$.
For simplicity, we place $c$ at some finite distance below  $\Ez$ (see Figure \ref{fig}).

The next step is to expand $\FockH$ as\footnote{For technical reasons, we split $\FockH=\FockHminus+\FockHplus$, where $\FockHplus := 0\oplus\left(c-\boldKz-\left(\frac{N-\Np}{N-1}\right) \boldKo -\frac{1}{N-1}\boldKf\right)$, and expand only $\FockHminus$. To keep the notation simple, we will ignore this subtlety for the sketch of the proof. All details can be found in \cite{spectrum}.}
\begin{eqnarray}\label{eqn:expansion:Hminus}
\FockH
&=&\sum\limits_{j=0}^{a}\lN^{\frac{j}{2}}\FockHj +\lN^{\frac{a+1}{2}}\FockR_a\,,
\end{eqnarray}
with  $\FockH_j$ as in \eqref{FockHj}. The remainders $\FockR_a$, which are essentially the remainders of the Taylor series expansion of the square roots in \eqref{eqn:FockHN}, can be bounded above by powers of the number operator.
Making use of the expansion \eqref{eqn:expansion:Hminus}, we expand the resolvent of $\FockH$ around the resolvent of $\FockHz$ and integrate along the contour $\gan$, which finally yields
\begin{eqnarray}
\P
&=& \sum\limits_{\l=0}^a \lN^\frac{\l}{2}\P_\l+\lN^\frac{a+1}{2} \left(\FockB_P(a)+\FockB_Q(a)\right)
\label{eqn:exp:P}
\end{eqnarray}
for $\P_\l$ as in \eqref{eqn:Pna} and where
\begin{equation}
\FockBPn(a)=\sum\limits_{\nu=0}^{a}\sum\limits_{m=1}^{a-\nu}\sum\limits_{\substack{\bj\in\N^m\\[2pt]|\bj|=a-\nu}}\frac{1}{2\pi\i}\goint\frac{\Pn}{z-\FockHminus}\,\FockRnu\ResHzz\FockHjo\ResHzz\mycdots\FockH_{j_m} \ResHzz\dz\,\label{B:P}
\end{equation}
and
\begin{equation}\begin{split}
\FockBQn(a)
\;=\;&\hspace{-0.1cm}\sum\limits_{\nu=0}^{a} 
\sum\limits_{m=1}^{a-\nu}
\sum\limits_{\substack{\bj\in\N^m\\[2pt]|\bj|=a-\nu}}
\sum\limits_{\l=0}^{m} 
\sum\limits_{\substack{\bk\in\{0,1\}^{m+1}\\|\bk|=\l}}
\frac{1}{2\pi\i}\goint
\frac{\Qn}{z-\tFockH}\FockRnu
\frac{\FockIn_{k_1}}{z-\FockHz}\FockHjo\mycdots\FockH_{j_m}\frac{\FockIn_{k_{m+1}}}{z-\FockHz}\dz \qquad\quad
\label{B:Q}
\end{split}\end{equation}
for $\FockInk=\Pz$ if $k=0$ and $\FockInk=\Qz$ if $k=1$.
To control the error terms, we estimate the operators $\FockH_j$ and $\FockR_\nu$ in terms of powers of $(\Np+1)$, prove a uniform bound on moments of the number operator with respect to $\Chi$, i.e., 
\begin{equation}\label{eqn:moments:number:op}
\lr{\Chi,(\Np+1)^b\Chi}\leq C(b)\,,
\end{equation}  
and control alternating products of number operators and resolvents of $\FockHz$ by means of the estimate
\begin{equation}
\Big\|(\Np+1)^{b+1}\ResHzz\bPhi\Big\| 
\leq C(b) \norm{(\Np+1)^{b}\bPhi} \,.
\end{equation}
To derive the expansion \eqref{eqn:expansion:E} of the ground state energy, we observe that
\begin{eqnarray}
\Tr\,\FockH\Pn & =& \frac{1}{2\pi\i}\Tr\goint\frac{\FockH}{z-\FockH}\dz \;=\;   \Ezn+\frac{1}{2\pi\i}\Tr\goint\frac{z-\Ezn}{z-\FockH}\dz
\end{eqnarray}
and derive from this the expansion
\begin{equation}\label{eqn:expansion:E}
\Tr\,\FockH\Pn 
=\Ezn+\sum\limits_{\l=1}^a\lN^\frac{\l}{2}\sum\limits_{\nu=1}^\l\sum\limits_{\substack{\bj\in\N^\nu\\|\bj|=\l}}\Tr\frac{1}{2\pi\i}\goint\ResHzz\FockHjo\ResHzz \mycdots\FockHjnu\frac{z-\Ezn}{z-\FockHz}\dz+\mathcal{O}(\lN^\frac{a+1}{2})\,.
\end{equation}
All half-integer powers of $\lN$  in \eqref{eqn:expansion:E} vanish by parity, which can be seen by conjugating with the  unitary map $\UP$ acting as 
$\UP\ad(f)\UP =-\ad(f)
$
(recall from \eqref{FockHj} that $\FockHj$ contains an even/odd number of creation/annihilation operators for $j$ even/odd). After some lengthy computations, this yields \eqref{def:E_l}.\\

\subsection{Proof of Corollary \ref{cor:RDM:static}}\label{subsec:proof:RDM}
To prove Corollary \ref{cor:RDM:static}, one first observes that $\gamma_N^{(1)}$ can be decomposed as
\begin{eqnarray}
\gamma_N^{(1)}\label{eqn:decomposition:gamma}
= \pp + \frac{1}{\sqrt{N}} \big(|\varphi\rangle \langle \beta_{\Chi}| + |\beta_{\Chi} \rangle \langle \varphi|\big) +\frac{1}{N} \Big(\gamma_\Chi- \pp\, \Tr\,\P \Np\Big),
\end{eqnarray}
where $\gamma_{\Chi}$ denotes the one-body reduced density matrix of $\Chi$ with kernel 
$\gamma_\Chi(x;y)=\langle\Chi,\ad_ya_x\Chi\rangle$
and where $\beta_\chi:\R^d\to\mathbb{C}$ is defined as
\begin{equation}\label{eqn:bChiNt}
\beta_\Chi(x) := \Tr\,\P \,\sqrt{1-\frac{\Np}{N}} \,a_x
\end{equation}
(see \cite[Section 3.5]{QF}).
Next, one expands the $N$-dependent expressions in \eqref{eqn:decomposition:gamma} in powers of $\lN^{1/2}$ and estimates the remainders using (a generalized version of) Proposition \ref{prop:static:technical}. We will show this for  $a=1$; the higher orders follow similarly using estimates from~\cite{spectrum}.\\

For $A\in\cL(\fH)$, \eqref{eqn:decomposition:gamma} yields
\begin{subequations}\label{eqn:gamma}
\begin{eqnarray}
&&\hspace{-1.5cm}\left|\Tr A\gamma_N^{(1)}-\Tr\, A\gamma_{1,0}-\lN\Tr\,A\gamma_{1,1}\right|\nonumber\\
&\leq&\left|\Tr\,\P\ad(A\varphi)\frac{\sqrt{N-\Np}}{N}-\lN\Tr\,\P_1\ad(A\varphi)\right|\label{eqn:gamma:1}\\
&&+\left|\Tr\,\P\frac{\sqrt{N-\Np}}{N}a(A\varphi)-\lN\Tr\,\P_1 a(A\varphi)\right|\label{eqn:gamma:2}\\
&&+\left|\frac{1}{N}\Tr\,\d\Gamma(A)\P-\lN\Tr\,\d\Gamma(A)\Pz\right|\label{eqn:gamma:3}\\
&&+\left|\lr{\varphi,A\varphi}\right|\left|\frac{1}{N}\Tr\,\P\Np-\lN\Tr\,\Pz\Np\right|\,.\label{eqn:gamma:5}
\end{eqnarray}
\end{subequations}
In the first line, we expand $\sqrt{N-\Np}/N=\lN^{1/2} + \lN^{3/2}\mathbb{R}$, where $\FockR$ is a function of $\Np$ such that $\norm{\FockR\bPhi}\ls\norm{(\Np+1)\bPhi}$ for any $\bPhi\in\Fock$ (see \cite[Section 5H, eqn.\ (5-64b)]{QF}). By parity, 
\begin{equation}
\Tr\,\Pz\ad(A\varphi)\FockR = \Tr\,\Pz\ad(A\varphi)=0\,,
\end{equation}
hence
\begin{eqnarray}
\eqref{eqn:gamma:1}
&\leq&\lN^\frac12\left|\Tr\,\P\ad(A\varphi)-\Tr\,(\Pz+\lN^\frac12\P_1)\ad(A\varphi)\right|\nonumber\\
&&+\lN^\frac32\left|\Tr\,\P\ad(A\varphi)\FockR-\Tr\,\Pz\ad(A\varphi)\FockR\right|\,\label{eqn:gamma:4}.
\end{eqnarray}
Since 
\begin{equation}
\norm{\ad(A\varphi)\FockR\bPhi}\leq\onorm{A}\norm{(\Np+1)^\frac32\bPhi}\,,\qquad
\norm{\ad(A\varphi)\bPhi}\leq\onorm{A}\norm{(\Np+1)^\frac12\bPhi}\,,
\end{equation}
one shows as in the proof of \cite[Theorem 1]{spectrum} that $\eqref{eqn:gamma:4}\ls \lN^2\onorm{A}$. The estimate of \eqref{eqn:gamma:2} works analogously.
For the third line in \eqref{eqn:gamma}, one notes that $|1/N-\lN|\ls\lN^2$ and that $\Tr\,\P_1\d\Gamma(A)=0$ by parity, hence
\begin{eqnarray}
\eqref{eqn:gamma:3}
&\ls&\lN^2\left|\Tr\,A\gamma_\Chi\right|+\lN\left|\Tr\,(\P-\Pz-\lN^\frac12\P_1)\d\Gamma(A)\right|
\;\ls\;\lN^2\onorm{A}
\end{eqnarray}
as above, where we used that $\norm{\d\Gamma(A)\bPhi}\leq\onorm{A}\norm{(\Np+1)\bPhi}$ for any $\bPhi\in\Fock$. Analogously, we derive the bound $\eqref{eqn:gamma:5}\ls\lN^2\onorm{A}$, making use of the fact that finite moments of $\Np$ with respect to $\Chiz$ and $\Chi$ are bounded uniformly in $N$ (\cite[Lemmas 4.7(d) and 5.6(a)]{spectrum}). This concludes the proof of Corollary \ref{cor:RDM:static} by duality of compact and trace class operators. \qed

\subsection{Extensions}\label{sec:spectrum:extensions} 

The results proven in \cite{spectrum} are more general than what we have presented so far. In this section, we briefly comment on some extensions of Theorem \ref{thm:eigenstates:spectrum}.

\subsubsection{Unbounded interaction potentials}

One extension  concerns  unbounded interaction potentials, including the three-dimensional repulsive Coulomb potential. In fact, we can replace Assumption \ref{ass:v} by the following assumption:\\

\noindent\textbf{Assumption 1'}. \textit{Let $v:\R^d\to\R$ be measurable with $v(-x)=v(x)$ and $v\not\equiv 0$, and assume that there exists a constant $C>0$ such that, in the sense of operators on $\cQ(-\Delta)=H^1(\R^d)$,
\begin{equation}
|v|^2\leq  C\left(1-\Delta\right)\,.  \label{eqn:ass:v:2:Delta:bound}
\end{equation}
Besides, assume that $v$ is of positive type.}\\

In this situation, we require one additional assumption, ensuring that the $N$-body state exhibits complete BEC with not too many particles outside the condensate:
\begin{assumption}\label{ass:cond}
Assume that there exist constants $C_1\geq0$, $0<C_2\leq 1$, and a function $\varepsilon:\N\to\R_0^+$ with 
$$\lim\limits_{N\to\infty} N^{-\frac13}\varepsilon(N) \leq C_1\,,$$
such that 
\begin{equation}\label{eqn:ass:cond}
\HN-N\eH\geq C_2 \sum\limits_{j=1}^N\hH_j-\varepsilon(N)
\end{equation}
in the sense of operators on $\D(\HN)$.
\end{assumption}
Under these more general assumptions, several new issues arise, at the core of which is the problem that $\d\Gamma(v)$ cannot be bounded by powers of $\Np+1$ alone. 
This affects the proof of Proposition \ref{prop:static:technical} at multiple points; most notably, it becomes considerably more difficult to obtain the uniform bound on moments of the number operator \eqref{eqn:moments:number:op}.

\subsubsection{Excited states}\label{subsec:ext:excitations}

The analysis in \cite{spectrum} extends to the low-energy eigenstates of $\HN$, i.e., it includes all eigenstates with an energy of order one above the ground state energy. 
In this situation, the expansion must be done more carefully, since the excited eigenvalues $\Ez^{(n)}>\Ez$ of $\FockHz$ can be degenerate, and the degeneracy of eigenvalues of $\FockHN$ may change in the limit $N\to\infty$. For instance,  an eigenvalue $\Ez^{(n)}$ of $\FockHz$ could be twice degenerate, with  two distinct eigenvalues $E^{(n_1)}_{\leq N}\neq E^{(n_2)}_{\leq N}$ of $\FockHN$ such that 
$$\lim\limits_{N\to\infty} E^{(n_1)}_{\leq N} =\Ez^{(n)} =\lim\limits_{N\to\infty} E^{(n_2)}_{\leq N}\,.$$
In this case, we expand the projector
\begin{equation}\label{eqn:Pn}
\P^{(n)}=\frac{1}{2\pi\i}\oint_{\gamma^{(n)}}\ResHz\dz
\end{equation}
around 
\begin{equation}
\Pz^{(n)}=\frac{1}{2\pi\i}\oint_{\gamma^{(n)}}\ResHzz\dz\,,
\end{equation}
where $\gamma^{(n)}$ is a $\mathcal{O}(1)$ contour around $\Ez^{(n)}$ with a finite distance to the remaining spectrum of $\FockHz$. Since $\gamma^{(n)}$ encloses both poles $E^{(n_1)}_{\leq N}$ and $E^{(n_2)}_{\leq N}$ of $(z-\FockH)^{-1}$, the contour integral \eqref{eqn:Pn} gives precisely the sum of the two spectral projectors of $\FockH$ corresponding to $E^{(n_1)}_{\leq N}$ and $E^{(n_2)}_{\leq N}$. 

In \cite{spectrum}, we show that there is a constant $C(a,n)$, which, in particular, depends on $|\Ez^{(n)}|$, such that 
\begin{equation}\label{eqn:ext:excitations}
\left|\Tr\,\FockA\P^{(n)}-\sum\limits_{\l=0}^a\lN^\frac{\l}{2}\Tr\,\FockA\P_\l^{(n)}\right|\leq C(a,n) \lN^\frac{a+1}{2} \onorm{\FockA}
\end{equation}
for sufficiently large $N$. The coefficients $\P_\l^{(n)}$  are defined analogously to $\P_\l$ from \eqref{eqn:Pna} but with $\Pz$ replaced by $\Pz^{(n)}$.
Note that the statement is non-trivial only for states with an energy of order one above the ground state energy because the constant $C(a,n)$ depends on $|\Ezn|$.\\

To state the generalization of the expansion \eqref{eqn:thm:spectrum} to the low-energy spectrum of $\HN$, we  need some more notation.
We denote by 
$$\EN\equiv\EN^{(0)}<\EN^{(1)}<\dots<\EN^{(\nu)}<\dots$$
the eigenvalues of $\HN$, and by $\dnu$ the degeneracy of $\EN^{(\nu)}$ (we follow the convention of counting eigenvalues without multiplicity). 
Given an eigenvalue $\Ez^{(n)}$ of $\FockHz$, we collect the indices $\nu$ of the eigenvalues $\ENnu$ that converge to $N\eH+\Ez^{(n)}$ for some given $n$ in the index set
\begin{equation}\label{In}
\In:=\left\{\nu\in\N_0:\, \lim\limits_{N\to\infty}\big(\ENnu-N\eH\big)=\Ezn^{(n)}\right\}\,.
\end{equation}
The generalization of \eqref{eqn:thm:spectrum} to excited eigenvalues $\EN^{(n)}$ is then given by
\begin{equation}\label{eqn:cor:expansion}
\left|{\sum\limits_{\;\nu\in\In}\dnu\ENnu} -  \dzn N\eH -  \sum\limits_{\l=0}^a\lN^\l\En_\l^{(n)} \right|\; \leq\; C(a,n) \lN^{a+1}\,,
\end{equation}
where $\dzn$ denotes the degeneracy of $\Ez^{(n)}$ and where $\En_\l^{(n)}$ is defined as in \eqref{def:E_l} but with $\Pz$ is replaced by $\Pz^{(n)}$. The constant $C(a,n)$ depends on $|\Ez^{(n)}|$.

\subsubsection{Expectation values of unbounded operators}
Finally, \cite{spectrum} yields an asymptotic expansion of expectation values of self-adjoint $m$-body operators $\Am$ which are relatively bounded with respect to $\sum_{j=1}^m(-\Delta_j+V(x_j))$,
i.e.,  
\begin{equation}\label{eqn:thm:A:unbounded}
\norm{\Am\psi}_{\fH^m}\leq \fC \Big\|\sum_{j=1}^m (-\Delta_j+V(x_j)+1)\psi\Big\|_{\fH^m} \qquad \text{ for }\;\psi\in \D\Big(\sum_{j=1}^m (-\Delta_j+V(x_j))\Big) \,.
\end{equation}
For $\cAm$ the symmetrized version of $\Am$, 
\begin{equation}
\cAm:=\binom{N}{m}^{-1}\sum\limits_{1\leq j_1<\dots<j_m\leq N} \Amj\,,
\end{equation}
we prove that there exists a constant $C(m,a)$ such that 
\begin{equation}\label{eqn:thm:exp:P}
\left|\lr{\PsiN,\cAm\PsiN}-\sum\limits_{\l=0}^a \lN^\frac{\l}{2}\Tr\left((\UNp\, \cAm \UNp^*\oplus 0)\P_\l^{(n)}\right)
 \right|\leq C(m,a) \lN^\frac{a+2}{2} 
\end{equation}
for sufficiently large $N$. The statement extends to excited states as explained in Section \ref{subsec:ext:excitations}.

The rate in \eqref{eqn:thm:exp:P} is by a factor $\lN^{1/2}$ better than the error estimate in Proposition \ref{prop:static:technical}. To see this, one considers the operator
$$
\FockAmred=\UNp\, \left(\cAm-\lr{\varphi^{\otimes N},\cAm\varphi^{\otimes N}}\right) \UNp^*\oplus 0\,,
$$ 
where we have subtracted the condensate expectation value of $\cAm$ (which is of order one). Because of this subtraction, one can show that $\FockAmred$ satisfies the estimate
\begin{equation}\label{eqn:A}
\norm{\FockAmred\bPhi}_\Fp 
\ls \lN^{\frac12}\,,\qquad \Phi\in\{\Chi,\Chiz\}\,,
\end{equation}
and  Proposition \ref{prop:static:technical} for $\FockAmred$ concludes the proof.

\section{Dynamics}\label{sec:dynamics}
In the remaining part of these notes, we study the dynamics generated by the Hamiltonian $\HNfree$ from \eqref{HNfree} and explain the expansions \eqref{eqn:thm:dynamics} and \eqref{eqn:cor:RDM:dynamics} of the time-evolved $N$-body wave function $\PsiNfree$ and of the reduced one-body density $\gamma_N^{\free,(1)}$. We drop the superscript  $^\free$  and use the superscript $^\trap$ wherever it applies.

\subsection{Framework}

We study the solutions $\PsiN(t)$ of the time-dependent $N$-body Schrödinger equation \eqref{SE} generated by the Hamiltonian $\HN$ from \eqref{HNfree}, which describes a system of $N$ interacting bosons  without external trapping potential. As initial state, we take
$$\PsiN(0)=\PsiNtrap\,,$$
where $\PsiNtrap$ is the ground state of $\HNtrap$.

\subsubsection{Condensate}
As explained above, $\PsiNtrap$ exhibits BEC in the Hartree minimizer $\varphi^\trap$, and it is well known that this property is preserved by the time evolution. More precisely, 
\begin{equation}\label{eqn:RDM:convergence}
\Tr\left|\gamma_N^{(1)}(t)-|\pt\rangle\langle\pt|\right|\leq \frac{C(t)}{N}
\end{equation}
(see, e.g., \cite{chen2011,mitrouskas2016}), where  $\pt$ is the solution of the Hartree equation,
\begin{equation}\label{hpt}
\i \partial_t \pt = \left(-\Delta + v*|\pt|^2 - \mpt\right)\pt
=: \hpt\pt,
\qquad \varphi(0)=\varphi^\trap\,,
\end{equation}
with phase factor $\mpt = \tfrac{1}{2} \int_{\R^d} \left(v*|\varphi(t)|^2\right)(x) |\varphi(t,x)|^2\dx$.
The solution of \eqref{hpt} in $H^1(\R^d)$ is unique and exists globally. We define the projectors $\ppt$ and $\qpt$ analogously to \eqref{def:pp}.

\subsubsection{Excitations}

Analogously to \eqref{eqn:decomposition:PsiN}, we decompose the time-evolved $N$-body state $\PsiN(t)$ into the condensate $\pt$ and excitations $\ChiN(t)$ from the condensate. The excitation vector $\ChiN(t)$ is an element of the (truncated) excitation Fock space $\FNpt\subset\Fpt\subset\Fock$ defined analogously to \eqref{Fock:space}. 
When restricted to  the time-dependent excitation Fock space $\Fpt$, the number operator $\Number$ on the (time-independent) Fock space $\Fock$ counts the number of excitations around the time-evolved condensate $\pt^{\otimes N}$.
As before, the relation between $\PsiN(t)$ and  $\ChiN(t)$ is given by the (now time-dependent) unitary map $\UNpt$ defined analogously to \eqref{map:U}, namely
\begin{equation}
\ChiN(t)=\UNpt\PsiN(t)\,.
\end{equation}
The evolution of the excitations is determined by the Schrödinger equation
\begin{equation}\label{eqn:SE:Fock}
\i\partial_t\ChiN(t)=\FockHNpt\ChiN(t)\,,\qquad \ChiN(0)=\mathfrak{U}_{N,\varphi^\trap}\PsiN^\trap
\end{equation}
on $\FNpt$, generated by the excitation Hamiltonian
\begin{equation}\label{H:Fock}
\FockHNpt=\i(\partial_t\UNpt)\UNpt^*+\UNpt\HN\UNpt^*\,.
\end{equation}
For  convenience, we write $\FockHNpt$ as restriction to $\FNpt$ of a Hamiltonian $\FockHpt$ on $\Fock$, 
which can be expressed, analogously to \eqref{eqn:FockHN}, in terms of $N$, $\Number$ and operators  $\mathbb{K}^{\pt}_j$, which are defined analogously\footnote{To obtain the time-dependent operators $\mathbb{K}^{\pt}_j$ from \eqref{eqn:K:notation}, one replaces $\varphi$ by $\pt$, $h$ by $\hpt$, $\mu$ by $\mpt$ and $K_1$ by $K^\pt_1=\qpt\tilde{K}^\pt\qpt$ with  $\tilde{K}^\pt(x_1;x_2)= \overline{\varphi(t,x_2)}v(x_1-x_2)\varphi(t,x_1)$.} to \eqref{eqn:K:notation}.
Expanding the $N$-dependent expressions in a Taylor series yields (formally) the power series
\begin{eqnarray}\label{eqn:expansion:HNpt}
\FockHpt
&=&\FockHopt+\sum\limits_{n\geq 1}\lN^{\frac{n}{2}}\FockHnpt\,,
\end{eqnarray}
with coefficients $\FockH^\pt_j$  analogously to \eqref{FockHj}.
Note that the operator $\FockHpt$ preserves the truncation of $\FN$, whereas this property is lost when truncating the  expansion  after finitely many terms.

\subsubsection{Bogoliubov approximation}

The leading order $\FockHopt$ in \eqref{eqn:expansion:HNpt} is the time-dependent Bogoliubov Hamiltonian, which generates the Bogoliubov time evolution
\begin{equation}\label{Bog:eqn}
\i\partial_t\Chiz(t)=\FockHopt\Chio(t)\,,\qquad
\Chio(0)=\Chiz^\trap\,.
\end{equation}
It is well known that the solution of \eqref{Bog:eqn} approximates the solution $\ChiN(t)$ of \eqref{eqn:SE:Fock} to leading order, i.e.,
\begin{equation}
\lim\limits_{N\to\infty}\norm{\ChiN(t)-\Chiz(t)}_{\FNpt}=0
\end{equation}
(see, e.g., \cite{lewin2015,nam2015}).
This is a very useful approximation because the time evolution generated by $\FockHopt$ acts as a Bogoliubov transformation $\BogUts$ on $\Fock$. 
This means a huge simplification compared with the full $N$-body dynamics because it essentially reduces the $N$-body problem to the problem of solving a $2\times2$ matrix differential equation:
the corresponding Bogoliubov map $\BogV(t,s)$ on $\fH\oplus\fH$ is determined by the differential equation
\begin{equation}\label{eqn:V(t,s):early}
\i\partial_t\BogV(t,s)=\mathcal{A}(t)\BogV(t,s)\,, \qquad
\BogV(s,s)=\id
\end{equation}
with
\begin{equation}\label{eqn:BogV(t,s)}
\BogV(t,s)=\begin{pmatrix}
U_{t,s} &\Vbar_{t,s} \\ V_{t,s} & \Ubar_{t,s} \end{pmatrix}\,,\qquad
\mathcal{A}(t)=\begin{pmatrix} 
\hpt+\Kopt& -\Ktpt \\ 
\overline{\Ktpt}& -\left(\hpt+\overline{\Kopt}\right) \end{pmatrix} \,.
\end{equation}
Since it is a Bogoliubov transformation, the Bogoliubov time evolution preserves quasi-freeness. Hence, $\Chiz(t)$ is uniquely determined by  its two-point functions,
\begin{equation}\label{eqn:gamma:alpha}
\gamma_{\Chio(t)}(x,y)=\lr{\Chio(t),\ad_y a_x\Chio(t)}_\Fock\,,\qquad
\alpha_{\Chio(t)}(x,y)=\lr{\Chio(t),a_x a_y\Chio(t)}_\Fock\,,
\end{equation}
which can be computed directly from the two-point functions of $\Chio(0)$ as
\begin{subequations}
\begin{eqnarray}
\gChiot(x,y)
&=& \Big(\Vbar_{t,0}\gChioz^T\Vbar^*_{t,0}+U_{t,0}\gChioz U^*_{t,0}-\Vbar_{t,0}\aChioz^*U_{t,0}^*
-U_{t,0}\aChioz\Vbar_{t,0}^*\Big)(x,y) \nonumber\\
&& +\left(\Vbar_{t,0}\Vbar_{t,0}^*\right)(x,y)\,,\label{computing_gamma_t} \\
\aChiot(x,y)
&=& \left(U_{t,0}\aChioz\Ubar_{t,0}^* +\Vbar_{t,0}\aChioz^* V_{t,0}^*
-U_{t,0}\gChioz V_{t,0}^* - \Vbar_{t,0}\gChioz^T\Ubar_{t,0}^*\right)(x,y) \nonumber\\
&& +\left(U_{t,0}V_{t,0}^*\right)(x,y) 
\,.\label{computing_alpha_t} 
\end{eqnarray}
\end{subequations}
Alternatively, one obtains $\gChiot$ and $\aChiot$ by solving the system of differential equations 
\begin{subequations}\label{eqn:PDE}
\begin{eqnarray}
\i\partial_t\gamma_{\Chio(t)}
&=& \left(\hpt+\Kopt\right)\gamma_{\Chio(t)}-\gamma_{\Chio(t)}\left(\hpt+\Kopt\right) \nonumber\\
&& +\Ktpt\alpha^*_{\Chio(t)}-\alpha_{\Chio(t)}\big(\Ktpt\big)^*,
\label{eqn:gamma:PDE} 
\\
\i\partial_t\alpha_{\Chio(t)}
&=&\left(\hpt+\Kopt\right)\alpha_{\Chio(t)}
 +\alpha_{\Chio(t)}\left(\hpt+\Kopt\right)^T \nonumber\\
&& +\Ktpt+\Ktpt\gamma_{\Chio(t)}^T+\gChiot\Ktpt\label{eqn:alpha:PDE}
\end{eqnarray}
\end{subequations}
(see \cite{grillakis2013,nam2015}).

\subsection{Expansion of the dynamics}
\subsubsection{Expansion of the time-evolved wave function}
With the formal ansatz
\begin{equation}\label{eqn:ansatz:ChiN(t)_later}
\ChiN(t)\oplus0=\sum\limits_{\l=0}^\infty \lN^{\frac{\l}{2}} \Chil(t)\,,
\end{equation}
the Schrödinger equation \eqref{eqn:SE:Fock} leads to the set of equations
\begin{eqnarray}\label{eqn:differential:form:Chil}
\i\partial_t\Chil(t)&=&\FockHopt\Chil(t)+\sum\limits_{n=1}^\l\FockHnpt\Chi_{\l-n}(t)\,.
\end{eqnarray}
Motivated by \eqref{eqn:differential:form:Chil}, we define iteratively
\begin{equation}\label{eqn:int:form:Chil}
\Chil(t): = \BogUtz\Chil(0)-\i\sum\limits_{n=1}^\l\,\int\limits_0^t\BogUts\, \FockHnps\,\Chi_{\l-n}(s)\ds \,, 
\end{equation}
where $\BogUts$ denotes the Bogoliubov time evolution, i.e., the Bogoliubov transformation corresponding to the solution $\BogV(t,s)$ of \eqref{eqn:V(t,s):early}. To prove Theorem \ref{thm:dynamics}, we show that these function $\Chil$ are the coefficients in an asymptotic expansion of $\ChiN$:
\begin{proposition}\label{prop:norm:approx:dynamics}
Let Assumption \ref{ass:v:bd} be satisfied, let $a\in\N_0$ and denote by $\ChiN(t)$ the solution of~\eqref{eqn:SE:Fock}. 
Then $\Chil(t)\in\Fpt$ and there exists a constant $C(a)$ such that
\begin{equation}\label{eqn:prop:norm:approx:dynamics}
\Big\|\ChiN(t)-\sum\limits_{\l=0}^a \lN^{\frac{\l}{2}}\Chil(t)\Big\|_{\FN}\leq \e^{C(a) t} \lN^{\frac{a+1}{2}}
\end{equation}
for all $t\in \R$ and sufficiently large $N$. 
\end{proposition}
The growth of the constant $C(a)$ in $a$ can be estimated as
\begin{equation}
C(a)\leq C a^2\ln a\,.
\end{equation}
We do not expect this to be optimal, especially since Borel summability was shown for a comparable expansion in~\cite{ginibre1980}.
As a consequence of Proposition \ref{prop:norm:approx:dynamics}, the coefficients $\PsiNl(t)$ of the expansion \eqref{eqn:thm:dynamics} of  $\PsiN(t)$ are given by
\begin{equation}\label{eqn:PsiN_l}
\PsiNl(t):=\sum\limits_{k=0}^N\pt^{\otimes(N-k)}\otimes_s\big(\chi_\l(t)\big)^{(k)}\,.
\end{equation}
The higher orders $\Chil(t)$ are completely determined by the solution $\Chio(t)$ of the Bogoliubov equation as
\begin{equation}\label{eqn:Chil:explicit:final}
\Chil(t)=\sum\limits_{\substack{0\leq n\leq 3\l\\ n+\l\text{ even}}}
\sum\limits_{\bj\in\{-1,1\}^{n}}
\int\dx^{(n)}
\mathfrak{C}^{(\bj)}_{\l,n}(t;x^{(n)})\,\,\asjo_{x_1}\,\mycdots\,a^{\sharp_{j_{n}}}_{x_{n}}
\Chio(t)\,,
\end{equation}
where we used the notation 
\begin{equation}
a_x^{\sharp_{-1}}:=a_x\,,\qquad a_x^{\sharp_{1}}:=\ad_x\,.
\end{equation}
The $N$-independent functions $\mathfrak{C}_{\l,n}^{(\bj)}$ 
are given in terms the matrix entries $U_{t,s}$ and $V_{t,s}$ of the solution $\BogV(t,s)$ of \eqref{eqn:V(t,s):early} and the initial data.
For example,
\begin{subequations}
\begin{eqnarray}
\mathfrak{C}_{1,1}^{(1)}(t)
&=& \left(U_{t,0}(U_0^\trap)^*-\Vbar_{t,0}(\overline{V_0^\trap})^*\right)\Theta_1^\trap\,,\\
\mathfrak{C}_{1,1}^{(-1)}(t)
&=& \left( V_{t,0}(U_0^\trap)^*-\Ubar_{t,0}(\overline{V_0^\trap})^*
\right)\Theta_1^\trap\,,
\end{eqnarray}
\end{subequations}
for $\Theta_1^\trap$ as in \eqref{eqn:Theta:1}. Here, $U_0^\mathrm{trap}$ and $V_0^\mathrm{trap}$ denote the matrix entries of the Bogoliubov map corresponding to the Bogoliubov transformation $\BogUz^\trap$ that diagonalizes $\FockHz^\trap$.
The coefficients $\mathfrak{C}_{\l,n}^{(\bj)}$ with larger indices are constructed from this in a systematic iterative procedure. Since the general formula is very long and not particularly insightful, we refrain from stating it here and refer to \cite[Eqn.\ (5.51)]{QF}.\\

The higher orders $\Chil(t)$ satisfy a generalized Wick rule for the ``mixed'' correlation functions 
\begin{equation}\label{eqn:mixed:corr}
\lrt{a^{\sharp_1}_{x_1}\cdots a^{\sharp_n}_{x_n}}_{\l,k}:=\lr{\Chil(t),a^{\sharp_1}_{x_1}\cdots a^{\sharp_n}_{x_n}\Chi_k(t)}\,.
\end{equation}
\newpage
\begin{proposition} \emph{(Generalized Wick Rule)}\label{prop:wick}
{
\begin{itemize}
\item \label{cor:gen:Wick:odd}
If $k+\l+n$ odd, 
\begin{equation}\label{nlk_correlations_odd}
\lrt{a^{\sharp_{j_1}}_{x_1}\cdots a^{\sharp_{j_n}}_{x_{n}}}_{\l,k} = 0 \,.
\end{equation}

\item \label{cor:gen:Wick:even}
If $k+\l+n$ even,
\begin{eqnarray}\label{nlk_correlations_even}
&&\hspace{-1cm}\lrt{a^{\sharp_{j_1}}_{x_1}\cdots a^{\sharp_{j_n}}_{x_{n}}}_{\l,k} \nonumber\\
&&\hspace{-0.8cm}=\sum\limits_{\substack{b=n\\\text{even}}}^{n+3(\l+k)} \sum\limits_{\bm\in\{-1,1\}^b}\,
\sum\limits_{\sigma\in P_{b}}\,\prod\limits_{i=1}^{b/2}
\int\dy^{(b)}\mathfrak{D}^{(\bj;\bm)}_{\l,k,n;b}(t;\xn;y^{(b)})\lrt{a_{y_{\sigma(2i-1)}}^{\sharp_{m_{\sigma(2i-1)}}}a_{y_{\sigma(2i)}}^{\sharp_{m_{\sigma(2i)}}}}_{0,0}\qquad
\end{eqnarray}
for $P_{b}$ the set of pairings defined in \eqref{pairings}. The functions $\mathfrak{D}^{(\bj;\bm)}_{\l,k,n;b}$ are determined by the coefficients $\fC$ from \eqref{eqn:Chil:explicit:final} (see \cite[Corollary 3.5]{QF} for the precise formula).
\end{itemize}
}
\end{proposition}

\subsubsection{Expansion of the one-body reduced density matrix}
As an application of \eqref{eqn:prop:norm:approx:dynamics}, we derive the expansion \eqref{eqn:cor:RDM:dynamics} of the one-body reduced density matrix. The coefficients $\gamma_{N,\l}^{(1)}$ in \eqref{eqn:cor:RDM:dynamics} are given by the trace class operators with kernels
\begin{subequations}\label{eqn:gamma:higher:orders}
\begin{eqnarray}
\gamma_{1,0}(t;x;y)&:=& \varphi(t,x)\overline{\varphi(t,y)}\,, \\
\gamma_{1,\l}(t;x;y)&:=& \sum\limits_{m=1}^{\l}\Bigg[
\sum\limits_{k=0}^{\l-m}\sum\limits_{n=0}^{2m-1}\tilde{c}_{\l-m,k}\,\bigg( \varphi(t,x)\lrt{ \ad_y(\Number-1)^k}_{n,2m-n-1}\nonumber\\
&&\hspace{4cm}+\lrt{(\Number-1)^ka_x}_{n,2m-n-1}\overline{\varphi(t,y)}
\bigg) \nonumber\\
&&+\sum\limits_{n=0}^{2m-2}\tilde{c}_{\l-m}\left(\lrt{\ad_y a_x}_{n,2m-n-2}
-\varphi(t,x)\overline{\varphi(t,y)}\lrt{\Number}_{n,2m-n-2}\right)\Bigg]\label{gamma_correction_a}\qquad
\end{eqnarray}
\end{subequations}
with $\tilde{c}_\l$ and $\tilde{c}_{\l,k}$ as in \eqref{def:c:tilde} and where we used the notation \eqref{eqn:mixed:corr}.
For example, the leading order  of the expansion is $\gamma_0^{(1)}(t)=\ppt$, which recovers \eqref{eqn:RDM:convergence}. The next-to-leading order is given by
\begin{equation}
\gamma_1^{(1)}(t)
=|\pt\rangle\langle\bzo(t)|
+|\bzo(t)\rangle\langle\pt|
+\gChiot-\Tr\,\gChiot\ppt\,,
\end{equation}
where the function $\bzo:\R^d\to\C$ is the solution of 
\begin{eqnarray}\label{intro_red_dens_corr}
\i\partial_t\bzo(t) 
&=&\left(\hpt+\Kopt\right)\bzo(t)+\Ktpt\overline{\bzo(t)} \nonumber\\
&& +\big(\Kthpt\big)^*\aChiot+\Tr_1\big(\Kthpt\gChiot\big)+\Tr_2\big(\Kthpt\gChiot\big)\,.
\end{eqnarray}
Here, $\gamma_{\Chio(t)}$ and $\alpha_{\Chio(t)}$ are the Bogoliubov two-point functions as in \eqref{eqn:gamma:alpha}, and we used the notation $\Tr_1A:=\int\dz A(z,\,\cdot\,;z)$ and $\Tr_2 A:=\int\dz A(\,\cdot\,,z;z) $, for an operator $A:\fH\to\fH^2$.

\subsection{Strategy of proof}
To prove Proposition \ref{prop:norm:approx:dynamics}, we first show that the functions $\Chil(t)$ defined in \eqref{eqn:int:form:Chil} are elements of $\Fpt$, by proving that
\begin{equation}\label{eqn:moments:Chil(t)}
\lr{\Chil(t),(\Number+1)^b\Chil(t)}_\Fock \ls \e^{C(\l,b)t}
\end{equation}
for any $b\in\N_0$. 
To this end, we re-write $\Chil(t)$ as 
\begin{eqnarray}
\Chil(t) &=& \BogUtz \Chil(0) \nonumber\\
&&+ \sum_{n=0}^{\l-1} \sum_{m=1}^{\l-n} \sum_{\substack{\bj \in \N^m \\ |\bj| = \l - n}} (-\i)^m \int\limits_0^t \ds_1 \int\limits_0^{s_1} \ds_{2}\, \mycdots\hspace{-5pt} \int\limits_0^{s_{m-1}} \ds_m \, \tilde{\mathbb{H}}^{(j_1)}_{t,s_1} \mycdots \,\tilde{\mathbb{H}}^{(j_m)}_{t,s_m} \, \BogUtz \Chi_n(0)\qquad\label{solution_chil_integral_form}
\end{eqnarray}
with
\begin{equation}\label{definition_of_Bog_trafo_Hns}
\tilde{\mathbb{H}}^{(n)}_{t,s} := \BogUts\FockH^\ps_n\BogUts^*\,,
\end{equation}
bound the operators $\tilde{\mathbb{H}}^{(n)}_{t,s}$ by powers of $(\Number+1)$, and make use of the fact that any finite moment of $\Number$ with respect to $\Chi_n(0)$ is bounded since $\Chi_n(0)=\Chi_n^\trap$ from \eqref{chi_def_thm}.
To prove \eqref{eqn:prop:norm:approx:dynamics}, we expand $\FockHpt$ in a Taylor series with remainder analogously to \eqref{eqn:expansion:Hminus}, prove an estimate the remainder in terms of $\Number$, and make use of \eqref{eqn:moments:Chil(t)} to close a Gronwall argument for the function $\tilde{\Chi}_a(t)=\ChiN(t)\oplus 0 -\sum_{\l=0}^a\lN^{\l/2}\Chil(t)$.

To prove Corollary \ref{cor:RDM:dynamics}, one decomposes $\gamma_N^{(1)}(t) $ analogously to \eqref{eqn:decomposition:gamma} and expands it in powers of $\lN^{1/2}$, which yields expressions containing correlation functions of $\ChiN$, 
\begin{equation}
\lrt{a^{\sharp_1}_{x_1}\cdots a^{\sharp_n}_{x_n}}_{N}:=\lr{\ChiN(t),a^{\sharp_1}_{x_1}\cdots a^{\sharp_n}_{x_n}\ChiN(t)}_{\FN}\,.
\end{equation}
Finally, we show that, in a suitable sense,
\begin{eqnarray}
\lrt{a^{\sharp_1}_{x_1}\cdots a^{\sharp_n}_{x_n}}_{N}
& =& \sum_{\ell=0}^{a} \lN^{\frac{\l}{2}} \sum_{m=0}^{\ell} \lr{\Chim(t) ,a^{\sharp_1}_{x_1}\cdots a^{\sharp_{n}}_{x_{n}} \Chi_{\ell-m}(t)}_\Fock + \mathcal{O}\Big(\lN^{\frac{a+1}{2}}\Big).
\end{eqnarray}
where all half-integer powers of $\lN$ vanish by the generalized Wick rule (Proposition \ref{prop:wick}).

\subsection{Extensions}
The results proven in \cite{QF} are more general than what was stated so far, namely they admit a larger class of initial data. It is not necessary to start the time evolution in the ground state $\PsiN^\trap$ of the trapped system (or in any low-energy eigenstate of $\HNtrap$), but it suffices if the initial state satisfies the following assumption:
\begin{assumption}\label{ass:initial:data}
Let $\tilde{a}\in\N_0$. Let $\PsiN(0)\in\mathcal{D}(\HN)$, define $\ChiN(0)=\UNpz\PsiN(0)$, and assume that there exists a constant $C(\tilde{a}) > 0$ such that
\begin{equation}\label{eqn:initial:expansion}
\left\|\ChiN(0)-\sum\limits_{\l=0}^{\tilde{a}} \lN^\frac{\l}{2}\Chil(0)\right\|_{\FN}\leq C(\tilde{a}) \, \lN^{\frac{\tilde{a}+1}{2}}\,,
\end{equation}
where the functions $\Chil(0)$ are defined as follows:
\begin{itemize}
\item  
Let $\tilde{\nu}\in\N_0$, let  $\BogUz$  be a Bogoliubov transformation on $\Fpz$, and let $\{f_j\}_{j=1}^{\tilde{\nu}}\subset\{\pz\}^\perp$ be some orthonormal system. Define
\begin{equation}\label{def:Chi0:0}
\Chi_0(0) : = \BogUz  a^{\dagger}\big(f_1 \big) \,\mycdots\, a^{\dagger}\big(f_{\tilde{\nu}}\big) \,|\Omega\rangle\,.
\end{equation}
\item 
For $1\leq\l\leq \tilde{a}$, let
\begin{equation}\label{eqn:Chil:0:normal:order}
\Chil(0)
=
\sum\limits_{\substack{0\leq m\leq 3\l\\m+\l\text{ even}}}\;\,
\sum\limits_{\mu=0}^m
\int\dx^{(\mu)}\dy^{(m-\mu)}
\tilde{\mathfrak{a}}^{(\l)}_{m,\mu}\big(x^{(\mu)};y^{(m-\mu)}\big)\ad_{x_1}\,\mycdots \ad_{x_\mu} a_{y_1}\,\mycdots a_{y_{m-\mu}}\, \Chi_{0}(0)\,,
\end{equation}
where $\mathfrak{a}^{(\l)}_{n,m,\mu}\big(x^{(\mu)};y^{(m-\mu)}\big)$ are the kernels of some $N$-independent bounded operators.
\end{itemize}
\end{assumption}
Moreover, our analysis generalizes to the case where $\Chio(0)$ is given as a linear combination of Bogoliubov transformed states with different particle numbers $\tilde{\nu}$. 
It is clear that this is satisfied by any superposition of low-energy eigenstates of $\HNtrap$.

\subsection{Related results}

We conclude with a brief overview of closely related results in the literature. 
The first derivation of higher order corrections is due to Ginibre and Velo \cite{ginibre1980,ginibre1980_2}, who consider the classical field limit $\hbar\to0$ of the dynamics generated by a Hamiltonian on Fock space with coherent states as initial data.
They construct a Dyson expansion of the unitary group $W(t,s)$ in terms of the time evolution generated by the Bogoliubov Hamiltonian; moreover, they prove that the expansion is Borel summable for bounded interaction potentials \cite{ginibre1980}.
The main difference to our work (apart from the Fock space setting) is that the authors expand the time evolution operator $W(t,s)$ in a perturbation series (and not the wave function). 
In contrast, we derive an expansion of the time-evolved wave function for a specific, physically relevant choice of initial data. This simplifies the approximation since fewer terms are required at a given order of the approximation because the state is expanded simultaneously with the Hamiltonian.

Another approach to higher order corrections in the mean-field regime in the $N$-body setting was proposed by Paul and Pulvirenti \cite{paul2019}. In that work, the authors approach the problem from a kinetic theory perspective and consider the dynamics of the reduced density matrices of the $N$-body state. Their approach is formally similar to ours, since Bogoliubov theory in the sense of linearization of the Hartree equation is used for the expansion and an $a$-dependent but $N$-independent number of operations is required for the construction. In comparison, the main advantage of our approach is that the coefficients $\Chil$ in our approximation are completely independent of $N$. \\

Finally, a similar result in the $N$-body setting was obtained in a joint work with N.\ Pavlovi\'c, P.\ Pickl and A.\ Soffer \cite{corr}. In this paper, we expand the $N$-body time evolution in a Dyson series comparable to \eqref{eqn:int:form:Chil} but with one crucial difference: instead of using the Bogoliubov time evolution, the expansion is in terms of an auxiliary time evolution $\tilde{U}_\varphi(t,s)$ on $\fH^N$, whose generator has a quadratic structure comparable to the Bogoliubov Hamiltonian (sometimes called particle number preserving Bogoliubov Hamiltonian).

Unfortunately, this auxiliary time evolution $\tilde{U}_\varphi(t,s)$ is a rather inaccessible object, which implicitly still depends on $N$.
In particular, it is not clear to what extent computations are less complex with respect to the time evolution $\tilde{U}_\varphi(t,s)$ than with respect to the full $N$-body problem.
This problem was the original motivation for the work \cite{QF}, where we modified the construction precisely such as to make the approximations completely $N$-independent and accessible to computations. Eventually, this also led to the paper \cite{spectrum}, which was partially intended as a rigorous motivation of the assumptions on the initial data in \cite{QF}.

\subsection*{Acknowledgements}
It is a pleasure to thank Nata\v{s}a Pavlovi\'c, Sören Petrat, Peter Pickl, Robert Seiringer and Avy Soffer for the collaboration on the works \cite{corr,QF,spectrum}.
Funding from the European Union’s Horizon 2020 research and innovation programme under the Marie Sk{\textl}odowska-Curie Grant Agreement No.\ 754411 is gratefully acknowledged.

\renewcommand{\bibname}{References}
\bibliographystyle{abbrv}
    \bibliography{bib_file}

\begin{thebibliography}{10}

\bibitem{corr}
L.~Bo{\ss}mann, N.~Pavlovi\'c, P.~Pickl, and A.~Soffer.
\newblock Higher order corrections to the mean-field description of the
  dynamics of interacting bosons.
\newblock {\em J.\ Stat.\ Phys.}, 178(6):1362--1396, 2020.

\bibitem{QF}
L.~Bo{\ss}mann, S.~Petrat, P.~Pickl, and A.~Soffer.
\newblock Beyond {Bogoliubov} dynamics.
\newblock {\em Pure Appl.\ Anal.}, 3(4):677--726, 2021.

\bibitem{spectrum}
L.~Bo{\ss}mann, S.~Petrat, and R.~Seiringer.
\newblock Asymptotic expansion of low-energy excitations for weakly interacting
  bosons.
\newblock {\em Forum Math.\ Sigma}, 9:e28, 2021.

\bibitem{chen2011}
L.~Chen, J.~O. Lee, and B.~Schlein.
\newblock Rate of convergence towards {Hartree} dynamics.
\newblock {\em J. Stat. Phys.}, 144(4):872--903, 2011.

\bibitem{ginibre1980_2}
J.~Ginibre and G.~Velo.
\newblock The classical field limit of non-relativistic bosons. {II.}
  {Asymptotic} expansions for general potentials.
\newblock {\em Ann. Inst. H. Poincar\'{e} Physique th{\'e}orique},
  33(4):363--394, 1980.

\bibitem{ginibre1980}
J.~Ginibre and G.~Velo.
\newblock The classical field limit of nonrelativistic bosons. {I.} {Borel}
  summability for bounded potentials.
\newblock {\em Ann. Phys.}, 128(2):243--285, 1980.

\bibitem{grech2013}
P.~Grech and R.~Seiringer.
\newblock The excitation spectrum for weakly interacting bosons in a trap.
\newblock {\em Commun. Math. Phys.}, 322(2):559--591, 2013.

\bibitem{grillakis2013}
M.~Grillakis and M.~Machedon.
\newblock Pair excitations and the mean field approximation of interacting
  bosons, {I}.
\newblock {\em Commun. Math. Phys.}, 324(2):601--636, 2013.

\bibitem{lewin2014}
M.~Lewin, P.~T. Nam, and N.~Rougerie.
\newblock Derivation of {H}artree's theory for generic mean-field {Bose}
  systems.
\newblock {\em Adv. Math.}, 254:570--621, 2014.

\bibitem{lewin2015}
M.~Lewin, P.~T. Nam, and B.~Schlein.
\newblock Fluctuations around {Hartree} states in the mean field regime.
\newblock {\em Amer. J. Math.}, 137(6):1613--1650, 2015.

\bibitem{lewin2015_2}
M.~Lewin, P.~T. Nam, S.~Serfaty, and J.~P. Solovej.
\newblock {B}ogoliubov spectrum of interacting {B}ose gases.
\newblock {\em Commun. Pure Appl. Math.}, 68(3):413--471, 2015.

\bibitem{mitrouskas2016}
D.~Mitrouskas, S.~Petrat, and P.~Pickl.
\newblock Bogoliubov corrections and trace norm convergence for the {Hartree}
  dynamics.
\newblock {\em Rev. Math. Phys.}, 31(8), 2019.

\bibitem{nam2015}
P.~T. Nam and M.~Napi{\'o}rkowski.
\newblock Bogoliubov correction to the mean-field dynamics of interacting
  bosons.
\newblock {\em Adv. Theor. Math. Phys.}, 21(3):683--738, 2017.

\bibitem{nam2020_2}
P.~T. Nam and M.~Napi{\'o}rkowski.
\newblock Two-term expansion of the ground state one-body density matrix of a
  mean-field {Bose} gas.
\newblock {\em Calc. Var. Partial Differential Equations}, 60(3):1--30, 2021.

\bibitem{paul2019}
T.~Paul and M.~Pulvirenti.
\newblock Asymptotic expansion of the mean-field approximation.
\newblock {\em Discrete Contin. Dyn. Syst. A}, 39(4):1891--1921, 2019.

\bibitem{pizzo2015}
A.~Pizzo.
\newblock Bose particles in a box {I}. {A} convergent expansion of the ground
  state of a three-modes {Bogoliubov Hamiltonian}.
\newblock {\em arXiv:1511.07022}, 2015.

\bibitem{pizzo2015_2}
A.~Pizzo.
\newblock Bose particles in a box {II}. {A} convergent expansion of the ground
  state of the {Bogoliubov Hamiltonian} in the mean field limiting regime.
\newblock {\em arXiv:1511.07025}, 2015.

\bibitem{pizzo2015_3}
A.~Pizzo.
\newblock Bose particles in a box {III}. {A} convergent expansion of the ground
  state of the {H}amiltonian in the mean field limiting regime.
\newblock {\em arXiv:1511.07026}, 2015.

\bibitem{seiringer2011}
R.~Seiringer.
\newblock The excitation spectrum for weakly interacting bosons.
\newblock {\em Commun. Math. Phys.}, 306(2):565--578, 2011.

\end{thebibliography}
\end{document}